\documentclass[envcountsame]{llncs-alternate}
\usepackage{graphicx} 
\usepackage{epsfig}
\usepackage{stmaryrd}
\usepackage{latexsym}
\usepackage{amssymb}
\usepackage{daytime}
\usepackage{eepic}
\usepackage{url}
\usepackage{verbatim}
\usepackage{wasysym}
\usepackage{cite}
\usepackage{comment}
\usepackage{fixtoc}
\def\thisPaperTitle{On Generalized Records and \\ Spatial Conjunction in Role Logic}

\usepackage{defs}
\title{\thisPaperTitle}

\author{Viktor Kuncak and Martin Rinard}
\institute{
        MIT Computer Science and Artificial Intelligence Laboratory\\
        {\tt $\{$vkuncak, rinard$\}$@csail.mit.edu} \\
        MIT CSAIL Technical Report No 942 \ \ {\tiny VK0104}
}

\begin{document}

\sloppy

\maketitle

\renewcommand{\thefootnote}{\fnsymbol{footnote}}
\footnotetext{Version of \today, \Daytime.}

\renewcommand{\thefootnote}{\arabic{footnote}}

\vspace*{-1em}
\begin{abstract}
We have previously introduced role logic as a notation for
describing properties of relational structures in shape
analysis, databases and knowledge bases.  A natural fragment
of role logic corresponds to two-variable logic with
counting and is therefore decidable.

We show how to use role logic to describe open and closed
records, as well the dual of records, inverse records.  We
observe that the spatial conjunction operation of separation
logic naturally models record concatenation.  Moreover, we
show how to eliminate the spatial conjunction of formulas of
quantifier depth one in first-order logic with counting.  As
a result, allowing spatial conjunction of formulas of
quantifier depth one preserves the decidability of
two-variable logic with counting.  This result applies to
two-variable role logic fragment as well.

The resulting logic smoothly integrates type system and
predicate calculus notation and can be viewed as a natural
generalization of the notation for constraints arising in
role analysis and similar shape analysis approaches.
\end{abstract}



\vspace*{-1em}
\smartparagraph{Keywords:}
Records, Shape Analysis, Static Analysis, Program Verification,
Two-Variable Logic with Counting, Description Logic, Types

\setcounter{tocdepth}{4}
\vspace*{-1em}
\tableofcontents

\section{Introduction}

In \cite{KuncakRinard03OnRoleLogic} we have introduced {\em
  role logic}, a notation for describing properties of
relational structures in shape analysis, databases and
knowledge bases.  Role logic notation aims to combine the
simplicity of role declarations
\cite{KuncakETAL02RoleAnalysis} and the well-established
first-order logic.
Role logic is closed under all boolean operations and
generalizes boolean shape analysis constraints
\cite{KuncakRinard03OnBooleanAlgebraShapeAnalysisConstraints}.
Role logic formulas easily translate into the traditional
first-order logic notation.  Despite this generality, role
logic enables the concise expression of common properties of
data structures in imperative programs that manipulate
complex data structures with mutable references.  
In \cite[Section
4]{KuncakRinard03OnRoleLogic} we have established the
decidability of the fragment $\RLtwo$ of role logic by
exhibiting a correspondence with two-variable logic with
counting $C^2$
\cite{GraedelETAL97TwoVariableLogicCountingDecidable,
  PacholskiETAL00ComplexityResultsFirstOrderTwo}.

\smartparagraph{Generalized records in role logic}
In this paper we  give a
systematic account of field and slot declarations of role
analysis \cite{KuncakETAL02RoleAnalysis} by introducing a set
of role logic shorthands that allows concise description of
records.  Our basic idea is to generalize types to
unary predicates on objects.  Some of the aspects of our
notion of records that indicate its generality are:
\begin{enumerate}
\item We allow building new records by taking the
  conjunction, disjunction, or negation of records.
\item In our notation, a record indicates a property of an
  object at a particular program point; objects can satisfy
  different record specifications at different program
  points.  As a result, our records can express typestate
  changes such as object initialization
  \cite{StromYemini86Typestate,
    StromYellin93ExtendingTypestate,
    FaehndrichLeino03DeclaringCheckingNonTypesObjectOrientedLanguage,
    DeLineFahndrich01EnforcingHighLevelProtocols,
    DeLineFahndrich04TypestateForObjects}
  and more general changes in relationships between objects
  such as movements of objects between data structures
  \cite{KuncakETAL02RoleAnalysis, KuncakETAL01RolesTechRep, SagivETAL02Parametric}.
\item We allow {\em inverse records} as a dual of records
  that specify incoming edges of an object in the graph of
  objects representing program heap.  Inverse records allow
  the specification of aliasing properties of objects,
  generalizing unique pointers.  Inverse records enable the
  convenient specification of movements of objects that
  participate in multiple data structures.
\item We allow the specification of both open and closed
  records.  Closed records specify a complete set of
  outgoing and incoming edges of an object.  Open records
  leave certain edges unspecified, which allows orthogonal
  data structures to be specified independently and then
  combined using logical conjunction.
\item We allow the concatenation of generalized records
  using a form of spatial conjunction of separation logic,
  while remaining within the decidable fragment of
  two-variable role logic.
\end{enumerate}

\smartparagraph{Separation logic} Separation logic
\cite{IshtiaqOHearn01BIAssertionLanguage,
  OHearnETAL01LocalReasoning,
  Reynolds00IntuitionisticReasoningMutable,
  Reynolds02SeparationLogic} is a promising approach for
specifying properties of programs in the presence of mutable
data structures.  One of the main uses of separation logic
in previous approaches is dealing with frame conditions
\cite{IshtiaqOHearn01BIAssertionLanguage,
  BirkedalETAL04LocalReasoningCopyingCollector}.  In
contrast, our paper identifies another use of spatial logic:
expressing record concatenation.  Although our approach is
based on essentially same logical operation of spatial
conjunction, our use of spatial conjunction for records is
more local, because it applies to the descriptions of the
neighborhood of an object.

To remain within the decidable fragment of role logic, we
give in Section~\ref{sec:spatialRLtwo} a construction that
eliminates spatial conjunction when it connects formulas of
quantifier depth one.  This construction also
illustrates that spatial conjunction is useful for reasoning
about counting stars
\cite{GraedelETAL97TwoVariableLogicCountingDecidable} of the
two-variable logic with counting $C^2$.  To our knowledge,
this is the first result that combines two-variable logic
with counting and a form of spatial conjunction.

\smartparagraph{Using the resulting logic} We can use
specifications written in our notation to describe
properties and relations between objects in programs with
dynamically allocated data structures.  These specifications
can act as assertions, preconditions, postconditions,
loop invariants or data structure invariants
\cite{KuncakETAL02RoleAnalysis, KuncakRinard03OnRoleLogic,
  LamETAL03OnModularPluggableAnalysesUsingSetInterfaces}.
By selecting a finite-height lattice of properties for a
given program fragment, abstract interpretation
\cite{CousotCousot77AbstractInterpretation} can be used to
synthesize properties of objects at intermediate program
points \cite{KuncakETAL02RoleAnalysis,
  SagivETAL02Parametric,
  RepsETAL03FiniteDifferencingLogicalFormulasStaticAnalysis,
  RepsETAL04SymbolicImplementationBestTransformer,
  YorshETAL04SymbolicallyComputingMostPrecise,
  YahavRamalingam04SeparationHeterogeneousAbstractions,  
  BallETAL01AutomaticPredicateAbstraction,  
  BallETAL02RelativeCompletenessAbstractionRefinement,
  HenzingerETAL04AbstractionsProofs}.  Decidability and
closure properties of our notation are essential for the
completeness and predictability of the resulting static
analysis
\cite{KuncakRinard04BooleanAlgebraShapeAnalysisConstraints}.

\smartparagraph{Contributions}
We summarize the main contributions of this paper as follows:
\begin{enumerate}
\item We present a logic which generalizes
  the concept of records in several directions
  (Section~\ref{sec:spatialRecords}).
  These generalizations are useful for expressing
  properties of objects and memory cells in imperative
  programs, and go beyond standard type systems.
\item We identify a novel use of separation logic:
  modelling the concatenation of generalized records.
\item We show how to translate role constraints from role
  analysis \cite{KuncakETAL02RoleAnalysis} to role logic
  (Section~\ref{sec:roleConstraints}).
\item We show that, under certain syntactic restrictions, we
  can translate spatial conjunction into other constructs of
  the decidable logic $\RLtwo$
  (Section~\ref{sec:spatialRLtwo}).  We therefore obtain a
  notation that extends $\RLtwo$ with a convenient way of
  describing record concatenation, and remains decidable.
 \item We present a translation of first-order logic with
   spatial conjunction and inductive definitions
   into second-order logic (Section~\ref{sec:fullSpatial}).
\end{enumerate}

\smartparagraph{Outline} Section~\ref{sec:roleLogic} reviews
the syntax and semantics of role logic.
Section~\ref{sec:spatialDef} defines spatial conjunction in
role logic and motivates its use for describing record
concatenation.  Section~\ref{sec:fieldExpr} and
Section~\ref{sec:spatialRecords} show how to use spatial
conjunction in role logic to describe a generalization of
records.  Section~\ref{sec:roleConstraints} demonstrates
that our notation is a generalization of the local
constraints arising in role analysis
\cite{KuncakETAL02RoleAnalysis} by giving a natural
embedding of role constraints into our notation.
Section~\ref{sec:spatialRLtwo} shows how to eliminate the
spatial conjunction connective $\spand$ from a spatial
conjunction $F_1 \spand F_2$ of two formulas $F_1$ and $F_2$
when $F_1$ and $F_2$ have no nested counting quantifiers;
this is the core technical result of this paper.  A
consequence of this is result is that we may allow certain
uses of spatial conjunction in $\RLtwo$ fragment of role
logic while preserving the decidability property of
$\RLtwo$.  Our extension of role logic with spatial
conjunction is therefore justified: it allows record-like
specifications to be expressed in a more natural way, and it
does not lead outside the decidable fragment.
Section~\ref{sec:remarks} contains remarks on preserving
the satisfiability of formulas in the presence of spatial
conjunction and shows how to encode the spatial
conjunction (with inductive definitions) in second-order
logic.  Section~\ref{sec:related} presents related work,
and Section~\ref{sec:conclusions} concludes.  Appendix
contains the details of the correctness proof for the
elimination of spatial conjunction from Section~\ref{sec:spatialRLtwo}.





\section{A Decidable Two-Variable Role Logic $\RLtwo$}
\label{sec:roleLogic}

\begin{figure}
  \vspace*{-3ex}
  \begin{equation*}
    \begin{array}{rcll}
      F & ::= & \multicolumn{2}{l}{
                A \mid f \mid \ID \mid F_1 \land F_2 \mid \lnot F \mid F' \mid \twid{F} \mid
                \card{{\geq}k} F} \mnl
      e & :: & \{ 1, 2 \} \to D \mnl
      \tr{A}e & = & \tr{A}(e\, 1) &
      \tr{f}e = \tr{f}(e\, 2, e\, 1) \mnl
      \tr{\ID}e & = & (e\, 2) = (e\, 1) \mnl
      \tr{F_1 \land F_2}e & = & (\tr{F_1}e) \land (\tr{F_2}e) &
      \tr{\lnot F}e = \lnot (\tr{F}e) \mnl
      \tr{F'}e & = & \tr{F} (e[1 \mapsto (e\, 2)]) \qquad &
      \tr{\twid{F}}e  = \tr{F} (e[1 \mapsto (e\, 2), 2 \mapsto (e\, 1)]) \mnl
      \tr{\card{{\geq}k} F}e & = &
        \multicolumn{2}{l}{| \{ d \in D \mid \tr{F} (e[1 \mapsto o, 2 \mapsto (e\, 1)]) \} | \geq k} \mnl
      F_1 \lor F_2 & \equiv & \lnot (\lnot F_1 \land \lnot F_2) \mnl
      F_1 \implies F_2 & \equiv & \lnot F_1 \lor F_2
    \end{array}
  \end{equation*}
  \caption{The Syntax and the Semantics of $\RLtwo$\label{fig:RLtwo}}
\end{figure}

Figure~\ref{fig:RLtwo} presents the two-variable role logic
$\RLtwo$ \cite{KuncakRinard03OnRoleLogic}.  We have proved
in \cite{KuncakRinard03OnRoleLogic} that $\RLtwo$ has the
same expressive power as two-variable logic with counting
$C^2$.  The logic $C^2$ is a first-order logic 1) extended with counting
quantifiers $\exists^{{\geq}k} x.F(x)$, saying that there are
at least $k$ elements $x$ satisfying formula $F(x)$ for
some constant $k$, and 2) restricted to allow only two
variable names $x$, $y$ in formulas.  An
example formula in two-variable logic with counting is
\begin{equationcl} \label{eqn:rlexample}
  \forall x. A(x) \implies (\forall y. f(x,y) \implies 
     \existseq{1}{x}{g(x,y)})
\end{equationcl}
The formula~(\ref{eqn:rlexample}) means that all nodes that
satisfy $A(x)$ point along the field $f$ to nodes that have
exactly one incoming $g$ edge.  Note that the variables $x$ and
$y$ may be reused via quantifier nesting, and that formulas
of the form $\existseq{k}{x}{F(x)}$ and
$\existsleq{k}{x}{F(x)}$ are expressible as boolean
combination of formulas of the form
$\existsgeq{k}{x}{F(x)}$.  The logic $C^2$ was shown
decidable in
\cite{GraedelETAL97TwoVariableLogicCountingDecidable} and
the complexity for the $C^2_1$ fragment of $C^2$ (with
counting up to one) was established in
\cite{PacholskiETAL00ComplexityResultsFirstOrderTwo}.  We
can view role logic as a variable-free version of $C^2$.
Variable-free logical notations are attractive as
generalizations of type systems because traditional type
systems are often variable-free.  The
formula~(\ref{eqn:rlexample}) can be written in role logic
as $[A \implies [f \implies \card{\geq 1} \twid{g}]]$ where
the construct $[F]$ is a shorthand for $\lnot \card{\geq 1}
\lnot F$ and corresponds to the universal quantifier.  The
expression $\twid{g}$ denotes the inverse of relation $g$.
This paper focuses on the use of role logic to describe
generalized records, see \cite{KuncakRinard03OnRoleLogic}
for further examples of using role logic and
\cite{Borgida95DescriptionLogicsDataManagement} for
advantages of variable-free notation in general.


\section{Spatial Conjunction}
\label{sec:spatialDef}

\begin{figure}
  \begin{equation*}
    \begin{array}{l}
      \tr{F_1 \spand F_2}e = \exists e_1,e_2.\
        \split\, e\, [e_1\, e_2]\ \land \ 
        \tr{F_1}e_1 \land \tr{F_2}e_2
      \\
      \split\, e\, [e_1\, e_2] = \mnls
      \quad
      \begin{array}[t]{l}
        \forall A \in \cala.\ \forall d \in D.\
        \begin{array}[t]{l}
          (e\, A)\, d \iff (e_1\, A)\, d \lor (e_2\, A)\, d \ \ \land \
          \lnot ((e_1\, A)\, d \land (e_2\, A)\, d) \ \ \land
        \end{array} \mnl
        \forall f \in \calf.\ \forall d_1, d_2 \in D. \mnl
        \quad
        \begin{array}[t]{l}
          (e\, f)\, d_1\, d_2 \iff (e_1\, f)\, d_1\, d_2 \lor (e_2\, f)\, d_1\, d_2 \ \ \land \
          \lnot ((e_1\, f)\, d_1\, d_2 \land (e_2\, f)\, d_1\, d_2) \mnls
        \end{array}
      \end{array} \\      
      \emp \ \equiv \ \squareb{\squareb{
          \bigwedge\limits_{A \in \cala} \lnot A \ \land\
          \bigwedge\limits_{f \in \calf} \lnot f}} \mnls
      \mbox{priority: $\land$ binds strongest, then $\spand$, then $\lor$} \mnls
      F \ssim G \ \ \mbox{means} \ \ \forall e.\, \tr{F}e=\tr{G}e \mnls
      (F_1 \spand F_2) \spand F_3 \ssim {F_1 \spand (F_2 \spand F_3)} \mnls
      F \spand \emp \ssim \emp \spand F \ssim F \mnls
      F_1 \spand F_2 \ssim F_2 \spand F_1 \mnls
      F_1 \spand (F_2 \lor F_3) \ssim F_1 \spand F_2 \, \lor\, F_1 \spand F_3
    \end{array}
  \end{equation*}
  \caption{Semantics and Properties of Spatial Conjunction $\spand$.
    \label{fig:spatialConjunction}}
\end{figure}

\noindent
Figure~\ref{fig:spatialConjunction} shows our semantics of
spatial conjunction $\spand$.  To
motivate our use of spatial conjunction, we first illustrate
how role logic supports the description of simple properties of
objects in a concise way.  Indeed, one of the design goals of
role logic is to have a logic-based specification language
where simple properties of objects are as convenient to
write as type declarations in a language like Java.
\begin{example} \label{exa:independent}
  The formula $\squareb{f \implies A}$ is true for an object
  whose every $f$-fields points to an $A$ object,
  $\squareb{g \implies B}$ means that every $g$-field points
  to a $B$ object, so
  \begin{equation*}
    \squareb{f \implies A} \land \squareb{g \implies B}
  \end{equation*}
  denotes the objects that has both $f$ pointing to an $A$
  object and $g$ pointing to a $B$ object.  Such specification
  is as concise as the following Java class declaration
\begin{verbatim}
class C { A f; B g; }
\end{verbatim}
\end{example}
Example~\ref{exa:independent} illustrates how the presence
of conjunction $\land$ in role logic enables combination of
orthogonal properties such as constraints on
distinct fields.  However, not all properties naturally
compose using conjunction.
\begin{example} \label{exa:needSpatial}
  Consider a program that contains three fields,
  modelled as binary relations $f$, $g$, $h$.
  The formula
  $
    P_f \ \equiv \ (\card{{=}1} f) \land (\card{{=}0} (g \lor h))
  $
  means that the object has only one outgoing $f$-edge
  and no other edges.
  The formula
  $
    P_g \ \equiv \ (\card{{=}1} g) \land (\card{{=}0} (f \lor h))
  $
  means that the object has only one outgoing $g$-edge
  and no other edges.  If we ``physically join'' two
  records, each of which has one field, we obtain 
  a record that has two fields, and is described by
  the formula
  \begin{equation*}
    P_{fg} \ \equiv \
    (\card{{=}1} f) \land (\card{{=}1} g) \land
    (\card{{=}0} h)
  \end{equation*}
  Note that it is {\em not} the case that $P_{fg} \ssim P_f
  \land P_g$.  More generally, no boolean combination of
  $P_f$ and $P_g$ yields $P_{fg}$.
\end{example}
Example~\ref{exa:needSpatial} prompts the question: is there
an operation that allows joining specifications that will
allow us to combine $P_f$ and $P_g$ into $P_{fg}$?  Moreover,
can we define such an operation on records viewed as arbitrary
formulas in role logic?

It turns out that there is a natural way to describe the set
of models of formula $P_{fg}$ in
Example~\ref{exa:needSpatial} as the result of ``physically
merging'' the edges (relations) of the models of $P_f$ and
models of $P_g$.  The merging of disjoint models of formulas
is the idea behind the definition of spatial conjunction
$\spand$ in Figure~\ref{fig:spatialConjunction}.  The
predicate $(\split\, e\, [e_1\, e_2])$ is true iff the
relations of the model (environment) $e$ can be split into
$e_1$ and $e_2$ and the notation generalizes to 
splitting into any number of environments.

\begin{example}
  For $P_f$, $P_g$, and $P_{fg}$ of
  Example~\ref{exa:needSpatial}, we have
$
    P_{fg} = P_f \spand P_g.
$
\end{example}

Note that the operation $\spand$ is associative and
commutative.  The formula $\emp$, which asserts that all
predicates are false, is the unit for $\spand$.  Moreover,
$\spand$ distributes over $\lor$.


\smartparagraph{A note on relationship with \cite{IshtiaqOHearn01BIAssertionLanguage}}
The semantics of spatial conjunction in
Figure~\ref{fig:spatialConjunction} match the semantics of
\cite{IshtiaqOHearn01BIAssertionLanguage}, with two
differences.

A small technical difference is that
Figure~\ref{fig:spatialConjunction} splits the edges of the
model (the tuples of the relations), whereas
\cite{IshtiaqOHearn01BIAssertionLanguage} splits the domain.
The difference arises because the elements of the domain in
\cite{IshtiaqOHearn01BIAssertionLanguage} are locations,
whereas the elements of our models are objects.  To
represent a location in our view, we would use a tuple
$\tu{o,f}$ where $o$ is an element of the domain and $f$ is
a field name.

A higher-level difference is that the use of spatial
logic we propose in this paper is the notation for records
(Section~\ref{sec:spatialRecords}), as opposed to the
description of global heap properties.  When used for
formulas of quantifier depth one
(Section~\ref{sec:spatialRLtwo}), spatial conjunction does
not even change the set of definable relations of
two-variable logic with counting.



\section{Field Complement}
\label{sec:fieldExpr}

As a step towards record calculus in role logic, this
section introduces the notion of a {\em field complement},
which makes it easier to describe records in role logic.
\begin{example} \label{exa:noComplement}
  Consider the formula $P_f \ \equiv \ (\card{{=}1} f) \land (\card{{=}0} (g \lor h))$
  from Example~\ref{exa:needSpatial}, stating the property
  that an object has only one outgoing $f$-edge and {\em no
    other edges}.  Property $P_f$ has little to do with $g$
  or $h$, yet $g$ and $h$ explicitly occur in $P_f$.
  Moreover, we need to know the entire set of relations in
  the language to write $P_f$; if the language
  contains an additional field $i$, the property $P_f$ would
  become $P_f \ \equiv \ (\card{{=}1} f) \land (\card{{=}0} (g \lor h \lor i))$.
  Note also that $\lnot f$ is not the same as $g \lor h \lor
  i$, because $\lnot f$ computes the complement of the value
  of the relation $f$ with respect to the universal set,
  whereas $g \lor h \lor i$ is the union of all relations
  other than $f$.
\end{example}
To address the notational problem illustrated in
Example~\ref{exa:noComplement}, we introduce the symbol
$\edges$, which denotes the union of all binary relations,
and the notation $\fc{f}$ ({\em field complement} of $f$),
which denotes the union of all relations other than $f$.
\begin{equationc}
  \begin{array}{lr}
    \edges   \equiv  \bigvee\limits_{g} g 
    & \qquad\qquad
    \fc{f}   \equiv \bigvee\limits_{g \neq f} g 
  \end{array}
\end{equationc}
This additional notation 
allows us to avoid explicitly listing all fields in the
language when stating properties like $P_f$.
\begin{example}
  Formula $P_f$ from Example~\ref{exa:noComplement} can be
  written as $P_f \ \equiv \ (\card{{=}1} f) \land
  (\card{{=}0} \fc{f})$, which mentions only $f$.  Even when
  the language is extended with additional relations, $P_f$
  still denotes the intended property.  Similarly, to denote
  the property of an object that has outgoing fields given
  by $P_f$ and has no incoming fields, we use the predicate
  $P_f \land \card{{=}0} \twid{\edges}$.
\end{example}
We use the notation $\edges$ and $\fc{f}$ to build the
notation for records and inverse records in
Section~\ref{sec:spatialRecords} below.

\smartparagraph{A note on ternary relation interpretation} It is 
possible to provide a notation for relations that
generalizes the notation $\edges$ and $\fc{f}$.  The idea of
this generalization is to change the definition of the
model (environment).  Instead of a model that specifies a binary
relation for each field, the model specifies the value of
one ternary relation $H$ and a unary tag-predicate for each
field name.  For example, instead of the model that provides
interpretations $f_I$ and $g_I$ for two binary relations $f$
and $g$, we could use the model that provides interpretation of
$\tr{H}$, where
\begin{equation*}
  \tr{H}o_1\, o_2\, n \ = \
  \begin{array}[t]{l}
    (n{=}f_0 \land f_I\, o_1\, o_2)\ \lor \mnl
    (n{=}g_0 \land f_I\, o_1\, o_2)
  \end{array}
\end{equation*}
and the interpretation of unary tag-predicates $f$ and $g$.
Here $f_0$ is an element of the domain that tags tuples
coming from $\tr{f}$, whereas $g_0$ tags tuples coming from
$\tr{g}$.  We interpret $f$ as a predicate that is true only
on the element $f_0$, and similarly $g$ as a predicate true
only on the element $g_0$.  We then introduce the
following dereferencing shorthand:
\begin{equation} \label{eqn:derefDef}
  \deref{F} \ \equiv \ \curlyb{H \land F}
\end{equation}
The expression $\deref{f}$ now denotes the original
interpretation of $f$, that is, $\tr{\deref{f}} = f_I$.
Moreover, $\deref{\lnot f}$ corresponds to field complement
$\fc{f}$, and $\deref{\boolTrue}$ corresponds to $\edges$.
Note that the expressions of the form $\deref{(\lnot f \land
  \lnot g)}$ are now also available.  Let $B$ be a boolean
combination of unary predicates denoting fields.  These unary
predicates are disjoint, so transforming $B$ into disjunctive
normal form and applying the property
\begin{equation*}
  \deref{(B_1 \lor B_2)} = \deref{B_1} \lor \deref{B_2}
\end{equation*}
which follows from~(\ref{eqn:derefDef}), allows transforming
$\deref{B}$ into a boolean combination of expressions of the
form $\deref{f}$ and $\deref{g}$.  This means that we 
obtain no additional expressive power using expressions of the
form $\deref{B}$ where $B$ is a boolean combination of unary
predicates denoting fields, so for simplicity we do not
consider such ``ternary relation interpretation'' further in
this paper.


\section{Records and Inverse Records}
\label{sec:spatialRecords}

In this section we use role logic with spatial conjunction
and field complement from Section~\ref{sec:fieldExpr} to
introduce a notation for records.  We also introduce inverse
records, which are dual to records, and correspond to slot
constraints in role analysis
\cite{KuncakETAL02RoleAnalysis}.

\begin{figure}  
  \begin{center}
    \begin{equation*}
      \begin{array}{rrcl}
        \mbox{multifield:} &
        f \pointstoMany A & \equiv & 
          \card{{=}0} (\fc{f} \lor \, (f \land \lnot A)) \mnl
        \mbox{field:} &
          f \pointstoWith{s} A & \equiv & 
          \begin{array}[t]{l}
            \card{s} (A \land f) \ \land \ f \pointstoMany A \\
            s \mbox{ of the form } {=}k, {\leq}k, \mbox{ or } {\geq}k, 
            \mbox{for } k \in \{ 0, 1, 2, \ldots \} \mnl
          \end{array} \\
        & f \pointsto A & \equiv & f \pointstoWith{{=}1} A \mnl
        \mbox{multislot:} &
        A \pointedbyMany f & \equiv & \card{{=}0} (\twid{\fc{f}} \lor \, (\twid{f} \land \lnot A)) \mnl
        \mbox{slot:} &
        A \pointedbyWith{s} f & \equiv & 
        \begin{array}[t]{l}
          \card{s} (A \land \twid{f}) \ \land \ A \pointedbyMany f \\
          s \mbox{ of the form } {=}k, {\leq}k, \mbox{ or } {\geq}k, 
            \mbox{for } k \in \{ 0, 1, 2, \ldots \} \mnl
        \end{array} \\
        & A \pointedby f & \equiv & A \pointedbyWith{{=}1} f
      \end{array}
    \end{equation*}
    \begin{equation*}
      \begin{array}{rcl}
        \fm & ::= & \field \mid \multifield \mnl
        \closedRecord & ::= & \fm \mid \closedRecord \spand \fm \mnl
        \openRecord & ::= & \closedRecord \spand \boolTrue \\
        \\
        \sm & ::= & \slot \mid \multislot \mnl
        \closedInverseRecord & ::= & \sm \mid \closedInverseRecord \spand \sm \mnl
        \openInverseRecord & ::= & \closedInverseRecord \spand \boolTrue
      \end{array}
    \end{equation*}
  \end{center}
  \caption{Record Notation\label{fig:records}}
\end{figure}

Figure~\ref{fig:records} presents the notation for records
and inverse records.  A \emph{field} predicate $f \pointsto A$ is
true for an object whose only outgoing edge in the graph
(model) is an $f$-edge terminating at $A$.  Dually, a \emph{slot}
predicate $A \pointedby f$ is true for an object whose only
incoming edge in the graph is an $f$-edge originating at
$A$.  A \emph{multifield} predicate $f \pointstoMany A$ is true iff
the object has any number of outgoing $f$-edges terminating
at $A$, and no other edges.  Dually, a \emph{multislot} predicate
$A \pointedbyMany f$ is true iff the object has any number
of incoming $f$-edges originating from $A$, and no other
edges.  We also allow notation $f \pointstoWith{s} A$ where
$s$ is an expression of the form ${=}k$, ${\leq}k$, or ${\geq}k$.  
This notation gives a bound on the number of outgoing
edges, and implies that there are no other outgoing edges.
We similarly introduce $A \pointedbyWith{s} f$.  A closed
record is a spatial conjunction of fields and multifields.
An open record is a spatial conjunction of a closed record
with $\boolTrue$.  While a closed record allows only the
listed fields, an open record allows any number of
additional fields.  Inverse records are dual to records, and
we similarly distinguish open and closed inverse records.

\begin{example}
  To describe a closed record whose only fields are $f$ and
  $g$ where $f$-fields point to objects in the set $A$ and
  $g$-fields point to objects in the set $B$, we use the
  predicate $ P_1 \ \equiv \ f \pointsto A \ \spand \ g
  \pointsto B.  $ The definition of $P_1$ lists all fields
  of the object.  To specify an open record which certainly
  has fields $f$ and $g$ but may or may not have other
  fields, we write $ P_2 \ \equiv \ f \pointsto A \ \spand \ 
  g \pointsto B \spand \boolTrue.  $ Neither $P_1$ nor $P_2$
  restrict incoming references of an object.  To specify
  that the only incoming references of an object are from
  the field $h$, we conjoin $P_2$ with the closed inverse
  record consisting of a single multislot $\boolTrue
  \pointedbyMany h$, yielding the predicate $ P_3 \ \equiv \ 
  P_2 \ \land \ \boolTrue \pointedbyMany h.  $ To specify
  that an object has exactly one incoming reference, and
  that the incoming reference is from the $h$ field and
  originates from an object belonging to the set $C$, we use
  $ P_4 \ \equiv \ P_2 \ \land \ C \pointedby h.  $ Note
  that specifications $P_3$ and $P_4$ go beyond most
  standard type systems in their ability to specify the
  incoming (in addition to the outgoing) references of
  objects.
\end{example}



\section{Role Constraints}
\label{sec:roleConstraints}

\begin{figure*}
\begin{equation*}
  \begin{array}{rcl}
    \trc{\q{fields}\ F;\ \q{slots}\ S;\ \q{identities}\ I;\
         \q{acyclic}\ A}
    &=& \begin{array}[t]{l}
        \trc{\q{fields}\ F}\ \land\ \trc{\q{slots}\ S}\ \land\ \mnl
        \tr{\q{identities}\ I}\ \land\ \tr{\q{acyclic}\ A}
        \end{array} \mnl
        \\
    \trc{\q{fields}\ f_1:S_1, \ldots, f_n:S_n}
    &=& f_1 \pointsto S_1\ \spand\ \ldots \ \spand\ f_n \pointsto S_n \mnl
    \trc{\q{slots}\ S_1.f_1,\ldots,S_n.f_n}
    &=& S_1 \pointedby f_1\ \spand\ \ldots\ \spand\ S_n \pointedby f_n \mnl
    \tr{\q{identities}\ f_1.g_1,\ldots,f_n.g_n}
    &=& \bigwedge_{i=1}^n [f_i \implies \twid{g_i}] \mnl
    \tr{\q{acyclic}\ f_1,\ldots,f_n}
    &=& \q{acyclic } (\bigvee_{i=1}^n f_i)
  \end{array}
\end{equation*}
\caption{Translation of Role Constraints \cite{KuncakETAL02RoleAnalysis}
into Role Logic Formulas
\label{fig:roleConstraintTransl}}
\end{figure*}

\begin{figure*}
\begin{equation*}
  \begin{array}{l}
    \tro{\q{fields}\ F;\ \q{slots}\ S;\ \q{identities}\ I;\
         \q{acyclic}\ A}
    = \begin{array}[t]{l}
        \tro{\q{fields}\ F}\ \land\ \tro{\q{slots}\ S}\ \land\ \mnl
        \tr{\q{identities}\ I}\ \land\ \tr{\q{acyclic}\ A} \mnl
        \end{array} \\
    \tro{\q{fields}\ f_1:S_1, \ldots, f_n:S_n}
    = \trc{\q{fields}\ f_1:S_1, \ldots, f_n:S_n} \spand \card{{=}0}{(\bigvee_{i=1}^n f_i)} \mnl
    \tro{g_1,\ldots, g_m\ \q{slots}\ S_1.f_1,\ldots,S_n.f_n}
    = \trc{\q{slots}\ S_1.f_1,\ldots,S_n.f_n} \spand \card{{=}0}{(\bigvee_{i=1}^m \twid{g_i})}
  \end{array}
\end{equation*}
\caption{Translation of Simultaneous Role Constraints 
\cite[Section 7.2]{KuncakETAL02RoleAnalysis}
into Role Logic Formulas. See also
Figure~\ref{fig:roleConstraintTransl}.
\label{fig:simultaneousRoleConstraintTransl}}
\end{figure*}

Role constraints were introduced in
\cite{KuncakETAL01LanguageRoleSpecifications, KuncakETAL02RoleAnalysis,
Kuncak01DesigningRoleAnalysis}.  In this section we show
that role logic is a natural generalization of role
constraints by giving a translation from role constraints to
role logic.  A logical view of role constraints is also
suggested in
\cite{KuncakRinard03ExistentialHeapAbstractionEntailment,
  KuncakRinard03ExistentialHeapAbstractionEntailment}.
A role is a set of objects that satisfy a
conjunction of the following four kinds of constraints:
field constraints, slot constraints, identities,
acyclicities.  In this paper we show that role logic naturally models
field constraints, slot constraints, and identities.
\footnote{Acyclicities go beyond first-order logic because they
involve non-local transitive closure properties.}

\smartparagraph{Roles describing complete sets of fields and
  slots} 
Figure~\ref{fig:roleConstraintTransl} shows the
translation of role constraints \cite[Section
3]{KuncakETAL02RoleAnalysis} into role logic formulas.
The simplicity of the translation
is a consequence of the notation for records that we have
developed in this paper.

\smartparagraph{Simultaneous Roles} In object-oriented programs, 
objects may participate in multiple data structures.  The
idea of simultaneous roles \cite[Section
7.2]{KuncakETAL02RoleAnalysis} is to associate one role for
the participation of an object in one data structure.  When
the object participates in multiple data structures, the
object plays multiple roles.  Role logic naturally models
simultaneous roles: each role is a unary predicate, and if
an object satisfies multiple roles, the the object satisfies
the conjunction of predicates.
Figure~\ref{fig:simultaneousRoleConstraintTransl} presents
the translation of field and slot constraints of
simultaneous roles into role logic.  Whereas the roles of
\cite[Section 3]{KuncakETAL02RoleAnalysis} translate to
closed records and closed inverse records, the simultaneous
roles of \cite[Section 7.2]{KuncakETAL02RoleAnalysis}
translate specifications that are closer to open records and
open inverse records.



\section{Eliminating Spatial Conjunction in $\RLtwo$}
\label{sec:spatialRLtwo}

\smartparagraph{Preserving the decidability} Previous
sections have demonstrated the usefulness of adding record
concatenation in the form of spatial conjunction to our
notation for generalized records.  However, a key question
remains: is the resulting extended notation decidable?  In
this section we give an affirmative answer to this question
by showing how to compute the spatial conjunction using the
remaining logical operations for a large class of record
specifications.

\smartparagraph{Approach} Consider two formulas $F_1$ and
$F_2$ in first-order logic with counting, where both $F_1$
and $F_2$ have quantifier depth one.  An equivalent way of
stating the condition on $F_1$ and $F_2$ is that there are
no nested occurrences of quantifiers.  (Note that we count
one application of $\existsgeq{k}{x}{P}$ as one quantifier,
regardless of the value $k$.)  We show that, under these
conditions, the spatial conjunction $F_1 \spand F_2$ can be
written as an equivalent formula $F_3$ where $F_3$ does not
contain the spatial conjunction operation $\spand$.  The
proof proceeds by writing formulas $F_1$, $F_2$ in a 
normal form, as a
disjunction of counting stars
\cite{GraedelETAL97TwoVariableLogicCountingDecidable}, and
showing that the spatial conjunction of counting stars is
equivalent to a disjunction of counting stars.

As a consequence of the results in this section, adding the
operation $\spand$ to logic with counting does not change
its expressive power provided that both $F_1$ and $F_2$
have quantifier depth at most one.  Here we allow $F_1$ and
$F_2$ themselves to contain spatial conjunction, because we
may eliminate spatial conjunction in $F_1$ and $F_2$
recursively.  Applying these results to two-variable logic
with counting $C^2$, we conclude that introducing into $C^2$ the spatial
conjunction of formulas of quantifier depth one 
preserves the decidability of $C^2$.  Furthermore, thanks to
the translations between $C^2$ and $\RLtwo$
in~\cite{KuncakRinard03OnRoleLogic}, if we allow the spatial
conjunction of $\RLtwo$ formulas with no nested $\card{}$ occurrences,
we preserve the decidability of the logic $\RLtwo$.  The formulas
of the resulting logic are given by
\begin{equationc}
  \begin{array}{rcl}
    F & ::= & 
      A \mid f \mid \ID \mid F_1 \land F_2 \mid \lnot F \mid F' \mid \twid{F} \mid
      \card{{\geq}k} F \mnls
      & \mid & F_1 \spand F_2, \mbox{ if $F_1$ and $F_2$ have no nested $\card{}$ occurrences}
  \end{array}
\end{equationc}
Note that record specifications in Figure~\ref{fig:records}
contain no nested $\card{}$ occurrences, so joining them
using $\spand$ yields formulas in the decidable fragment.
Hence, in addition to quantifiers and boolean operations,
the resulting logic supports a generalization of record
concatenation, and is still decidable; this decidability
property is what we show in the sequel.  We present the
sketch of the proof, see Appendix for proof details..

\subsection{Atomic Type Formulas}

In this section we introduce classes of formulas
that correspond to the model-theoretic notion of atomic type
\cite[Page 20]{Otto97BoundedVariableLogicsCounting} (see
\cite[Page 42]{Hodges93ModelTheory} and \cite[Page
78]{ChangKeisler90ModelTheory} for the notion of type in
general).  We then introduce formulas that describe the
notion of counting stars
\cite{GraedelETAL97TwoVariableLogicCountingDecidable,
  PacholskiETAL00ComplexityResultsFirstOrderTwo}.  We
conclude this section with 
Proposition~\ref{prop:depthOneNF}, which gives the normal
form for formulas of quantifier depth one.

If $\calc = C_1,\ldots,C_m$ is a finite set of formulas,
then a {\em cube over $\calc$} is a conjunction of the
form $C_1^{\alpha_1} \land \ldots C_m^{\alpha_m}$
where $\alpha_i \in \{0,1\}$, $C^1 = C$ and $C^0 = \lnot C$.
For simplicity, fix a finite language $L = \cala \cup \calf$
with $\cala$ a finite set of unary predicate symbols and $\calf$ a finite set of binary
predicate symbols.
We work in predicate
calculus with equality, and assume that the equality
``$=$'', where ${=} \notin \calf$, is present as a binary
relation symbol, unless explicitly stated otherwise.  We use $D$ to denote
a finite domain of interpretation and $e$
to denote a model with variable assignment; $e$ maps $\cala$ to $2^D$, maps
$\calf$ to $2^{D \times D}$ and maps variables to elements of $D$.
Let $x_1,\ldots,x_n$ be a finite list of distinct
variables.  Let $\calc$ be the set of all atomic formulas
$F$ such that $\FV{F} \subseteq \{ x_1,\ldots,x_n \}$.
The set $\calc$ is finite (in our case it has $|\cala|n +
(|\calf|+1)n^2$ elements).  We call a cube over $\calc$ a
\emph{complete atomic type (CAT) formula}.
\begin{example}
If $\cala = \{A\}$ and $\calf = \{f\}$,
then
\begin{equationc}
  \begin{array}{l}
    A(x_1) \land \lnot A(x_2)\ \land \mnl
    \lnot f(x_1,x_1) \land \lnot f(x_2,x_2) \land
    f(x_1,x_2) \land \lnot f(x_2,x_1) \ \land \mnl
    x_1 = x_1 \land x_2 = x_2 \land x_1 \neq x_2 \land x_2 \neq x_1
  \end{array}
\end{equationc}
is a CAT formula.
\end{example}
We may treat conjunction of literals as the set of literals,
so we say that ``a literal belongs to the conjunction'' and
apply set-theoretic operations on conjunctions of literals.

From the disjunctive normal form theorem for
propositional logic, we obtain the following
Proposition~\ref{prop:qfreeToCAT}.
\begin{proposition} \label{prop:qfreeToCAT}
  Every quantifier-free formula $F$ such that $\FV{F}
  \subseteq \{ x_1,\ldots,x_n \}$ is equivalent to a
  disjunction of CAT formulas $C$ such that $\FV{C} = \{
  x_1,\ldots,x_n \}$.
\end{proposition}

%
A CAT formula may be contradictory if, for example, it
contains the literal $x_i \neq x_i$ as a conjunct.  We next
define classes of CAT formulas that are satisfiable in the
presence of equality.
  Let $x_1,\ldots,x_n$ be distinct variables.  A
  \emph{general-case CAT (GCCAT)} formula is a CAT formula
  $F$ such that the following two conditions hold:
  1) $\FV{F} = \{ x_1,\ldots,x_n \}$;
  2) for all $1 \leq i,j \leq n$, the conjunct $x_i =
    x_j$ is in $F$ iff $i \equiv j$.
  Let $x_1,\ldots,x_n$ and $y_1,\ldots,y_m$ be distinct
  variables.  An {\em equality CAT (EQCAT)} formula
  is a formula of the form 
  $
    \bigwedge_{j=1}^m y_j = x_{i_j} \ \land\  F,
  $
  where $1 \leq i_1,\ldots,i_m
  \leq n$ and $F$ is a GCCAT formula such that $\FV{F} = \{
  x_1,\ldots,x_n \}$.
\begin{lemma}
  \label{lemma:equalingCAT}
  Every CAT formula $F$ is either contradictory, or is
  equivalent to an EQCAT formula $F'$ such that $\FV{F'} =
  \FV{F}$.
\end{lemma}
From Proposition~\ref{prop:qfreeToCAT} and
Lemma~\ref{lemma:equalingCAT}, we obtain the following
Proposition~\ref{prop:qfreeToEQCAT}.
\begin{proposition} \label{prop:qfreeToEQCAT}  
  Every quantifier-free formula $F$ such that $\FV{F}
  \subseteq \{ x_1,\ldots, x_n \}$ can be written as a
  disjunction of EQCAT formulas $C$ such that $\FV{C} = \{
  x_1,\ldots,x_n \}$.
\end{proposition}


\noindent
We next introduce the notion of an extension of a GCCAT
formula.  Let $x, x_1, \ldots, x_n$ be distinct variables
  and $F$ be a GCCAT formula such that
  $\FV{F} = \{ x_1,\ldots, x_n \}$.  We say that
  $F'$ is an $x$-extension of $F$, and write
  $F' \in \exts{F}{x}$ iff all of the following
  conditions hold:
  1) $F \land F'$ is a GCCAT formula;
  2) $\FV{F \land F'} = \{ x, x_1, \ldots, x_n \}$;
  3) $F$ and $F'$ have no common atomic formulas.
Note that if $\FV{F_1}=\FV{F_2}$, then
$\exts{F_1}{x}=\exts{F_2}{x}$ i.e.\ the set of extensions of
a GCCAT formula depends only on the free variables of the
formula; we introduce additional notation
$\exts{x_1,\ldots,x_n}{x}$ to denote $\exts{F}{x}$ for
$\FV{F}=\{x_1,\ldots,x_n\}$.



To define a normal form for formulas of quantifier depth
one, we introduce the notion of $k$-counting star.
If $p \geq 2$
is a non-negative integer, let $p^{+}$ be a new symbol which
represents the co-finite set of integers $\{ p, p+1 ,\ldots \}$.  Let
$C_p = \{ 0, 1, \ldots, p{-}1, p^{+} \}$.
If $c \in C_p$, by $\cexists{i}{x}{P}$ we mean
$\existseq{i}{x}{P}$ if $i$ is an integer, and
$\existsgeq{p}{x}{P}$ if $i=p^{+}$.  We say that a formula
$F$ has a {\em counting degree} of at most $p$ iff the only
counting quantifiers in $F$ are of the form
$\cexists{c}{x}{G}$ for some $c \in C_{p+1}$.
\begin{definition}[Counting Star Formula]
  \label{def:countingStar}
  Let $x$, $x_1,\ldots,x_n$, and $y_1,\ldots,y_m$ be
  distinct variables, $k \geq 1$ a
  positive integer, and $F$ a GCCAT formula such that
  $\FV{F} = \{ x_1,\ldots,x_n \}$.
  A \emph{$k$-counting star function for $F$}
  is a function $\gamma : \exts{F}{x} \to C_{k+1}$.
  A \emph{$k$-counting-star formula for $\gamma$} is a
  formula of the form
    \begin{equationc}
      \bigwedge_{j=1}^m y_j = x_{i_j} \ \land \ F \ \land
      \bigwedge_{F' \in \exts{F}{x}} \cexists{\gamma(F')}{x}{F'}
    \end{equationc}
    where $1 \leq i_1,\ldots,i_m \leq n$.
\end{definition}
Note that in Definition~\ref{def:countingStar}, formula
$\bigwedge_{j=1}^m y_j = x_{i_j} \land F$ is an EQCAT
formula, and formula 
$\bigwedge_{j=1}^m y_j = x_{i_j} \land F \land F'$ is an EQCAT 
formula for each $F' \in \exts{F}{x}$.

The following Proposition~\ref{prop:depthOneNF} shows that
formulas of quantifier depth at most one are equivalent to
disjunctions of counting stars.
\begin{proposition}[Depth-One Normal Form]
  \label{prop:depthOneNF}
  Let $F$ be a formula of such that $F$ has quantifier depth
  at most one, $F$ has counting degree at most $k$, and
  $\FV{F} \subseteq \{ x_1,\ldots,x_n \}$.  Then $F$ is
  equivalent to a disjunction of $k$-counting-star formulas
  $F_C$ where $\FV{F_C} = \{ x_1,\ldots, x_n \}$.
\end{proposition}

\subsection{Spatial Conjunction of Stars}
\label{sec:spatConjStars}

\smartparagraph{Sketch of the construction} 
Let $F_1$ and $F_2$ be two formulas of quantifier depth at
most one, and not containing the logical operation $\spand$.
By Proposition~\ref{prop:depthOneNF}, let $F_1$ be
equivalent to the disjunction of counting star formulas
$\bigvee_{i=1}^{n_1} C_{1,i}$ and let $F_2$ be equivalent to
the disjunction of counting star formulas
$\bigvee_{j=1}^{n_2} C_{2,j}$.  By distributivity of law of
$\spand$ with respect to $\lor$, we have
\begin{equationc}
    F_1 \spand F_2  \ssim 
      (\bigvee\limits_{i=1}^{n_1} C_{1,i}) \spand
      (\bigvee\limits_{j=1}^{n_2} C_{2,j})
      \ \ssim \
      \bigvee\limits_{i=1}^{n_1} \bigvee\limits_{j=1}^{n_2}
      C_{1,i} \spand C_{2,j}
\end{equationc}
In the sequel we show that a spatial conjunction of
counting-star formulas is either contradictory or is
equivalent to a disjunction of counting star formulas.  This
suffices to eliminate spatial conjunction of formulas of
quantifier depth at most one.  Moreover, if $F$ is any
formula of quantifier depth at most one, possibly containing
$\spand$, by repeated elimination of the innermost $\spand$
we obtain a formula without $\spand$.

To compute the spatial conjunction of counting stars we
establish an alternative syntactic form for counting star
formulas.  The idea of this alternative form is roughly to
replace a counting quantifier such as $\existseq{k}{x}{F'}$
with a spatial conjunction of $k$ formulas each of which has
the meaning similar to $\existseq{1}{x}{F'}$, and then
combine a formula $\existseq{1}{x}{F_1'}$ resulting from one
counting star with a formula $\existseq{1}{x}{F_2'}$
resulting from another counting star into the formula
$\existseq{1}{x}{(F_1' \ispand F_2')}$ where $\ispand$
denotes merging of GCCAT formulas by taking the union of
their positive literals.  We next develop this idea in greater detail.

\smartparagraph{Notation for spatial representation of stars}
  Let $\emptyGCCAT{x_1,\ldots,x_n}$ be the unique GCCAT formula
  $F$ with $\FV{F} = \{ x_1,\ldots,x_n \}$ such that the
  only positive literals in $F$ are literals $x_i = x_i$ for
  $1 \leq i \leq n$.
  Similarly, there is a unique
  formula $F' \in \exts{x_1,\ldots,x_n}{x}$ such that every
  atomic formula in $F'$ distinct from for $x=x$ 
  occurs in a negated literal.  We call $F'$ an
  \emph{empty extension} and denote it
  $\emptyExtension{x_1,\ldots,x_n}{x}$.

To compute a spatial conjunction of formulas $C_1$ and $C_2$
in the language $L$, we temporarily consider formulas in an
extended language $L' = L \cup \{ B_1, B_2 \}$ where $B_1$
and $B_2$ are two new unary predicates used to mark
formulas.  We use $B_1$ to mark formulas
derived from $C_1$, and use $B_2$ to mark formulas derived
from $C_2$.  
  For $m \in \{ \emptyset, \{1\}, \{2\}, \{1,2\} \}$, define
  \begin{equationc}
    \begin{array}{l@{\ \ \ }l}
      \Mark{\emptyset}{x} = \lnot B_1(x) \land \lnot B_2(x) & 
      \Mark{1}{x} = B_1(x) \land \lnot B_2(x) \\
      \Mark{2}{x} = \lnot B_1(x) \land B_2(x) &
      \Mark{1,2}{x} = B_1(x) \land B_2(x)
    \end{array}
  \end{equationc}
Note that, when we say that $F$ is a GCCAT formula, we mean
that $F$ is GCCAT formula in language $L$ (and thus $F$
mentions symbols only from $L$), even when we use $F$ as a
subformula of a larger formula in language $L'$.  Similarly,
expressions $\exts{x_1,\ldots,x_n}{x}$,
$\emptyExtension{F}{x}$, and $\emptyGCCAT{x_1,\ldots,x_n}$
all denote formulas in language $L$.  

On the other hand, $\allemptyExtension{F}{x}$ and $\empe$
are formulas in language $L'$.  Formula
$\allemptyExtension{F}{x}$ is an empty extension of $F$ in
language $L'$.  Formula $\empe$ asserts that
$x_1,\ldots,x_n$ have an empty GCCAT formula and that the
remaining elements have empty extension in $L'$.  Formula
$\empe$ does not constrain the values $B_1(x_i)$ and
$B_2(x_i)$, these values turn out to be
irrelevant.

\noindent
  Let $F' \in \exts{x_1,\ldots,x_n}{x}$.  Define
  \begin{equationc}
    \begin{array}{l}
      \allemptyExtension{x_1,\ldots,x_n}{x} \equiv
      \emptyExtension{x_1,\ldots,x_n}{x} \land \Mark{\emptyset}{x} \mnl
      \empe(x_1,\ldots,x_n)  \equiv
      \emptyGCCAT{x_1,\ldots,x_n} \ \land \
      \forall x.\,  (\bigwedge_{i=1}^n x \neq x_i) \implies 
      \allemptyExtension{x_1,\ldots,x_n}{x}  
    \end{array} 
  \end{equationc}
We write $\allemptyExtension{F}{x}$ for
$\allemptyExtension{x_1,\ldots,x_n}{x}$ if $\FV{F} =
\{x_1,\ldots,x_n\}$, and similarly for $\empe(F,x)$.  We write
simply $\empe$ if $F$ and $x$ are understood.

We next introduce formulas $\anyNumi{F'}{m}$ and
$\oneNumi{F'}{m}$, which are the building blocks for
representing counting star formulas.  Formula $\anyNumi{F'}{m}$
means that $F'$ marked with $m$ and
$\allemptyExtension{F}{x}$ are the only extensions of $F$ that
hold in the neighborhood of $x_1,\ldots,x_n$ ($F'$ may hold 
for \emph{any number} of neighbors).  Formula
$\oneNumi{F'}{m}$ means that
$F'$ holds for \emph{exactly one} element in the neighborhood of
$x_1,\ldots,x_n$, and all other neighbors have empty extensions.
More precisely,
  let $F' \in \exts{x_1,\ldots,x_n}{x}$.  Define
  \begin{equationc}
    \begin{array}{rcl}
      \anyNumi{F'}{m} & \equiv &
      \begin{array}[t]{@{}l}
        \emptyGCCAT{x_1,\ldots,x_n} \ \land \
        \forall x.\,  (\bigwedge_{i=1}^n x \neq x_i) \implies 
         (F' \land \Mark{m}{x}) \lor \allemptyExtension{F}{x}  \mnl
      \end{array} \\
      \oneNumi{F'}{m} & \equiv & \anyNumi{F'}{m}\ \land \
       \existseq{1}{x}{\ \ \bigwedge_{i=1}^n x \neq x_i\ \land\ 
       F' \land \Mark{m}{x}}
    \end{array}
  \end{equationc}
  where $m \in \{ \emptyset, \{1\}, \{2\}, \{1,2\} \}$.
Observe that $G \spand \empe \ssim G$ if $G \equiv
\anyNumi{F'}{m}$ or $G \equiv \oneNumi{F'}{m}$ for some $F'$
and $m$.  Also note that $\anyNumi{F'}{m} \spand \anyNumi{F'}{m} \sim
\anyNumi{F'}{m}$.

\begin{figure}[htbp]
  \vspace*{-3ex}
  \begin{equationc}
    \begin{array}{lcl}
      \multicolumn{3}{l}{
        \begin{array}{ccl}
          E \land F & - & \mbox{EQCAT formula} \\
          F & - & \mbox{GCCAT formula} \mnl
      \end{array}} \\
      \multicolumn{3}{l}{
      \str{m}{E \land F \land \exists^{s_1} x.F'_1 \land \ldots \land
        \exists^{s_k} x. F'_k} \ =} \\
    \multicolumn{3}{r}{
      = \ E \ \land\ \kstr{F} \spand \xstr{m}{\exists^{s_1} x.F'_1} \spand \ldots
      \spand \xstr{m}{\exists^{s_k} x.F'_k}} \mnl
    \multicolumn{3}{r}{
    \kstr{F} \ = \ F \ \land \ (\forall x.\, 
        (\bigwedge_{i=1}^n x \neq x_i) \implies \allemptyExtension{F}{x})} \mnl
    \xstr{m}{\exists^0 x.\, F'} & = & \empe \mnl
    \xstr{m}{\exists^{i+1} x.\, F'} & = & 
      \oneNumi{F'}{m} \spand \xstr{m}{\exists^i x.\, F'} \mnl
    \xstr{m}{\exists^{i^{+}} x.\, F'} & = &
      \xstr{m}{\exists^{i}\, x.\, F'} \spand 
      \anyNumi{F'}{m}
    \end{array}
  \end{equationc}
  \caption{Translation of Counting Stars to Spatial Notation
  \label{fig:translationToSpatial}}
\end{figure}

\smartparagraph{Translation of counting stars}
Figure~\ref{fig:translationToSpatial} presents the
translation of counting stars to spatial notation.  The idea
of the translation is to replace $\existseq{k}{x}{F'}$ with
the spatial conjunction of $k$ formulas $\oneNumi{F'}{m}
\spand \ldots \spand \oneNumi{F'}{m}$ where $m \in \{ \{1\},
\{2\} \}$.  The purpose of the marker $m$ is to ensure that
each of the $k$ witnesses for $x$ that are guaranteed to
exist by $\oneNumi{F'}{m} \spand \ldots \spand
\oneNumi{F'}{m}$ are distinct.  The reason that the
witnesses are distinct for $m \neq \emptyset$ is that no two
of them can satisfy $B_i(x)$ at the same time for $i \in m$.

To show the correctness of the translation in
Figure~\ref{fig:translationToSpatial}, define $e^m$ to be
the $L'$-environment obtained by extending $L$-environment
$e$ according to marking $m$, and $\eproj{e_1}$ to be the
restriction of an $L'$ environment $e_1$ to language $L$.
More precisely, if $e$ is an environment in language $L$, for $m \in \{
  \emptyset, \{1\}, \{2\}, \{1,2\} \}$, define environment
  $e^m$ in language $L'$ by 
1) $e^m\, r = e\, r$ for $r \in L$ and 2) for $q \in \{ 1, 2 \}$, let
$(e\, B_q)\, d = \boolTrue \iff 
        q \in m \ \land \ d \notin \{ e\, x_1, \ldots, e\, x_n\}$.
  Conversely, if $e_1$ is an environment in language $L'$,
  define environment $\eproj{e_1}$ in language $L$ by
  $\eproj{e_1}\, r = e_1\, r$ for all $r \in L$.
%
Lemma~\ref{lemma:translationCorrectness} below gives the
correctness criterion for translation in
Figure~\ref{fig:translationToSpatial}.  
\begin{lemma} \label{lemma:translationCorrectness}
  If $e$ is an environment for language $L$, $C$ a
  counting star formula in language $L$, and $m \in \{ \{ 1
  \}, \{ 2 \}, \{ 1,2 \} \}$, then $\tr{C}e =
  \str{m}{C}e^m$.
\end{lemma}

\begin{figure}
  \vspace*{-3ex}
  \begin{equationc}
    \begin{array}{rl}
      (1) & \oneNumi{T_1}{1} \spand \oneNumi{T_2}{2} \combto 
            \oneNumi{T_1 \ispand T_2}{1,2} \\
      (2) & \oneNumi{T_1}{1} \spand \anyNumi{T_2}{2} \combto 
            \oneNumi{T_1 \ispand T_2}{1,2} \spand \anyNumi{T_2}{2} \\
      (3) & \anyNumi{T_1}{1} \spand \oneNumi{T_2}{2} \combto
            \anyNumi{T_1}{1} \spand \oneNumi{T_1 \ispand T_2}{1,2} \\
      (4) & \anyNumi{T_1}{1} \spand \anyNumi{T_2}{2} \combto
            \anyNumi{T_1}{1} \spand  \anyNumi{T_2}{2} \spand \anyNumi{T_1 \ispand T_2}{1,2} \\
      (5) & \anyNumi{T}{1} \combto \empe \\
      (6) & \anyNumi{T}{2} \combto \empe
    \end{array} 
  \end{equationc}
  \caption{Transformation Rules for Combining 
    Spatial Conjuncts\label{fig:transRules}}
\end{figure}

\smartparagraph{Combining quantifier-free formulas}
Let $C_1 \spand C_2$ be a spatial conjunction of two
counting-star formulas
\begin{equationc}
  \begin{array}{l}
    C_1 \equiv E \land F_1 \land \exists^{s_{1,1}} x.F'_{1,1} \land \ldots \land
    \exists^{s_{1,k}} x. F'_{1,k} \\
    C_2 \equiv E \land F_2 \land \exists^{s_{2,1}} x.F'_{2,1} \land \ldots \land
    \exists^{s_{2,k}} x. F'_{2,l} \\
  \end{array}
\end{equationc}
where $F_1$ and $F_2$ are GCCAT formulas with $\FV{F_1} =
\FV{F_2} = \{x_1,\ldots,x_n\}$, $E \land F_1$ and $E \land
F_2$ are EQCAT formulas, and $E \equiv \bigwedge_{j=1}^m y_j
= x_{i_j}$.  

Note that we assume that the two GCCAT formulas
$F_1$ and $F_2$ have same free variables and that the
equalities $E$ in the two EQCAT formulas are the same.  This
assumption is justified because either 1) $C_1 \spand C_2$
make inconsistent assumptions about equalities among
$x_1,\ldots,x_n$, and therefore $C_1 \spand C_2$ is
equivalent to $\boolFalse$, or 2) $C_1 \spand C_2$ make same
assumptions about equalities among $x_1,\ldots,x_n$, so we
can rewrite $C_1$ and $C_2$ to satisfy the our assumption by
exchanging variables $x_i$ and $y_j$ in the definition of an
EQCAT formula.

To show how to transform formula $\str{1}{C_1} \spand
\str{2}{C_2}$ into a disjunction of formulas of the form
$\str{1,2}{C_3}$, we introduce the following notation.  If
$T$ is a formula, let $S(T)$ denote the set of positive
literals in $T_1$ that do not contain equality.  Let $T_1
\in \exts{F_1}{x}$ and $T_2 \in \exts{F_2}{x}$.  (Note that
$\exts{F_1}{x}=\exts{F_2}{x}$.)  We define the partial
operation $T_1 \ispand T_2$ as follows.  The result of $T_1
\ispand T_2$ is defined iff $S(T_1) \cap S(T_2) =
\emptyset$.  If $S(T_1) \cap S(T_2) = \emptyset$, then $T_1
\ispand T_2 = T$ where $T$ is the unique element of
$\exts{F_1}{x}$ such that $S(T) = S(T_1) \cup S(T_2)$.
Similarly to $\ispand$, we define the partial operation $F_1
\kispand F_2$ for $F_1$ and $F_2$ GCCAT formulas with
$\FV{F_1} = \FV{F_2} = \{ x_1,\ldots,x_n \}$.  The result of
$F_1 \kispand F_2$ is defined iff $S(F_1) \cap S(F_2) =
\emptyset$.  If $S(F_1) \cap S(F_2) = \emptyset$, then $F_1
\kispand F_2$ is the unique GCCAT formula $F$ such that
$\FV{F} = \{ x_1,\ldots,x_n \}$ and $S(F) = S(F_1) \cup
S(F_2)$.
The following Lemma~\ref{lemma:combiningQFree} notes that
$\ispand$ and $\kispand$ are sound rules for computing
spatial conjunction of certain quantifier-free formulas.
\begin{lemma} \label{lemma:combiningQFree}
  If $T_1, T_2 \in \exts{x_1,\ldots,x_n}{x}$
  then 
  $
    T_1 \spand T_2 \ \ssim\ T_1 \ispand T_2.
  $
  If $F_1$ and $F_2$ are GCCAT formulas
  with $\FV{F_1} = \FV{F_2} = \{ x_1,\ldots, x_n \}$,
  then
  $
    F_1 \spand F_2 \ \ssim\ F_1 \kispand F_2.
  $
\end{lemma}


\smartparagraph{Rules for transforming spatial conjuncts}
We transform formula $\str{1}{C_1} \spand
\str{2}{C_2}$ into a disjunction of formulas of the form
$\str{1,2}{C_3}$ as follows.

The first step in transforming $C_1 \spand C_2$ is to
replace $\kstr{F_1} \spand \kstr{F_2}$ with
$\kstr{F_1 \kispand F_2}$ if $F_1 \kispand F_2$ is
defined, or $\boolFalse$ if $F_1 \kispand F_2$ is not
defined.

The second step is summarized in
Figure~\ref{fig:transRules}, which presents rules for
combining conjuncts resulting from
$\xstr{1}{\exists^{s_1}.F_1}$ and $\xstr{2}{\exists^{s_2}
  x.F_2}$ into conjuncts of the form $\xstr{1,2}{\exists^s
  x.F}$.  The intuition is that $\anyNumi{T}{m}$ and
$\oneNumi{T}{m}$ represent a finite abstraction of all
possible neighborhoods of $x_1,\ldots,x_n$, and the
rules in Figure~\ref{fig:transRules} represent
the ways in which different portions of the neighborhoods
combine using spatial conjunction.  We apply the rules in
Figure~\ref{fig:transRules} modulo commutativity and
associativity of $\spand$, the fact that $\emp$ is
a unit for $\spand$, as well as the idempotence of $\anyNumi{T}{m}$.
Rules
$(1){-}(4)$ are applicable only when the occurrence of $T_1
\ispand T_2$ on the right-hand side of the rule is defined.
We apply rules $(1){-}(4)$ as long as possible, and then
apply rules $(5),(6)$.  Moreover, we only allow the sequences
of rule applications that eliminate all occurrences of
$\oneNumi{T}{1}$, $\anyNumi{T}{1}$,
$\oneNumi{T}{2}$, $\anyNumi{T}{2}$, leaving only
$\oneNumi{T}{1,2}$ and $\anyNumi{T}{1,2}$.  Note also that
the are only finitely many non-equivalent expressions that
can be obtained by sequences of applications of rules 
in Figure~\ref{fig:transRules}.  Namely, an application of
rules $(1)$--$(3)$ decreases the total number of spatial conjuncts
of the form $\oneNumi{T}{1}$ and $\oneNumi{T}{2}$, multiple applications
of rule $(4)$ to the same pair of spatial conjuncts are unnecessary
because of the idempotence of $\anyNumi{T_1 \ispand T_2}{1,2}$ (so we never perform them),
and rules $(5)$, $(6)$ reduce the total number of spatial conjuncts.
The following Lemma~\ref{lemma:rulesSound} gives partial
correctness of rules in Figure~\ref{fig:transRules}.
\begin{lemma} \label{lemma:rulesSound}
  If $G_1 \combto G_2$, then $G_2 \implies G_1$ is valid.
\end{lemma}
  Define $G_1 \yields G_2$ to hold iff both
  of the following two conditions hold:
  {\bf 1)} $G_2$ results from $G_1$ by replacing $\kstr{F_1}
    \spand \kstr{F_2}$ with $\kstr{F_1 \kispand F_2}$ if
    $F_1 \kispand F_2$ is defined, or $\boolFalse$ if $F_1
    \kispand F_2$ is not defined, and then applying some
    sequence of rules in Figure~\ref{fig:transRules} such
    that rules $(5),(6)$ are applied only when rules
    $(1){-}(4)$ are not applicable;
  {\bf 2)} $G_2$ contains only spatial conjuncts of the form
    $\oneNumi{T}{1,2}$ and $\anyNumi{T}{1,2}$.
From Lemma~\ref{lemma:rulesSound} and Lemma~\ref{lemma:combiningQFree}
we immediately obtain Lemma~\ref{lemma:yieldSound}.
\begin{lemma} \label{lemma:yieldSound}
  If $G_1 \yields G_2$, then $G_2 \implies G_1$ is valid.
\end{lemma}
The rule for computing the
spatial conjunction of counting star formulas is the following.
If $C_1$, $C_2$, and $C_3$ are counting star
formulas, 
define $\ERes{C_1}{C_2}{C_3}$ to hold iff
  $\str{1}{C_1} \spand \str{2}{C_2} \yields \str{1,2}{C_3}$.
We compute spatial conjunction by replacing $C_1 \spand C_2$
with $\bigvee_{\ERes{C_1}{C_2}{C_3}} C_3$.
Our goal is therefore to show the equivalence
\begin{equationcl} \label{eqn:computingSpatial}
  C_1 \spand C_2 \ \ssim\ \bigvee_{\ERes{C_1}{C_2}{C_3}} C_3
\end{equationcl}
The validity of $\bigvee_{\ERes{C_1}{C_2}{C_3}} C_3 \implies
(C_1 \spand C_2)$ follows from
Lemma~\ref{lemma:yieldSound} and
Lemma~\ref{lemma:translationCorrectness}.
\begin{lemma} \label{lemma:easyDirection}
  $(\bigvee_{\ERes{C_1}{C_2}{C_3}} C_3) \implies (C_1 \spand
  C_2)$ is a valid formula for every pair of counting star
  formulas $C_1$ and $C_2$.
\end{lemma}
We next consider the converse claim.  If $\tr{C_1 \spand C_2}e$, then
there are $e_1$ and $e_2$ such that $\split\, e\, e_1\,
e_2$, $\tr{C_1}e_1$, and $\tr{C_2}e_2$.  By considering the atomic
types induced in $e$, $e_1$ and $e_2$ 
by elements in $D \setminus \{ e\, x_1,\ldots,e\, x_n\}$,
we construct a sequence of $\combto$ transformations in Figure~\ref{fig:transRules} that
convert $\str{1}{C_1} \spand \str{2}{C_2}$ into a formula
$\str{1,2}{C_3}$ such that $\tr{C_3}e =\boolTrue$.
\begin{lemma} \label{lemma:difficultDirection}
  $C_1 \spand C_2 \implies \bigvee_{\ERes{C_1}{C_2}{C_3}} C_3$ 
  is a valid formula for every pair of counting star
  formulas $C_1$ and $C_2$.
\end{lemma}

From Lemma~\ref{lemma:easyDirection} and
Lemma~\ref{lemma:difficultDirection} we obtain the desired
Theorem~\ref{thm:equivalenceHolds}, which shows the
correctness of our rules for computing spatial conjunction
of formulas of quantifier depth at most one.
\begin{theorem} \label{thm:equivalenceHolds}
  The equivalence~(\ref{eqn:computingSpatial}) holds for every
  pair of counting star formulas $C_1$ and $C_2$.
\end{theorem}



\section{Further Remarks}
\label{sec:remarks}

In this section we present two additional remarks regarding
spatial conjunction.  The first remark notes that we must be
careful when extracting a subformula from a formula and
labelling it with a new predicate.  The second remark shows
how to encode spatial conjunction in second-order logic, thus
providing some insight into the expressive power of spatial
conjunction.

\subsection{Extracting Subformulas in the Presence of $\spand$}

In two-variable logic with counting $C^2$ we may efficiently
transform formula into an unnested form by introducing new
predicate names and naming subformulas using these
predicates.  This transformations is a standard step in
decidability proofs for two-variable logic with counting
\cite{GraedelETAL97TwoVariableLogicCountingDecidable,
  PacholskiETAL00ComplexityResultsFirstOrderTwo}.

The satisfiability of the resulting formula is
equivalent to the satisfiability of the original formula.
An extraction of a subformula $G$ and its replacement with a
new predicate $P$ can be justified by a substitution lemma of
the form:
\begin{equation*}  
  \tr{F[P:=G]}e = \tr{F}(e[P:=\tr{G}e])
\end{equation*}
where $e$ is the environment (model).  This substitution
lemma does not hold in the presence of spatial conjunction
that splits the values of newly introduced predicates.
Namely,
\begin{equation*}
  \tr{(F_1 \spand F_2)[P:=G]}e \ \implies\
  \tr{F_1 \spand F_2} (e[P:=\tr{G}e])
\end{equation*}
holds, but the converse implication does not hold because
the value $\tr{G}e$ of the relation $P$ might be split on
the right-hand side.

It is therefore interesting to divide predicates into
\emph{splittable} and \emph{non-splittable} predicates, and
have spatial conjunction split only the interpretations of
splittable predicates.  The substitution lemma then holds
when $P$ is a non-splittable predicate.

Note, however, that in the presence of non-splittable
predicates we cannot translate counting stars into spatial
notation and thus use unnested form to eliminate all spatial
conjunctions from first-order formulas.  As a
result, adding spatial conjunction of formulas of large
quantifier depth to two-variable logic with counting may
increase the expressive power of the resulting logic.

We also remark that if the language contains only one
splittable unary predicate $A_S$, then it is easy to
simulate the splitting of objects of the universe, which is
the semantics of spatial conjunction in
\cite{IshtiaqOHearn01BIAssertionLanguage}.  Namely, we use
some fixed unary predicate $A_0$ to denote all ``live''
objects, and make all quantifiers range only over the
objects that satisfy $A_0$.

\subsection{Representing $\spand$ in Second-Order Logic}
\label{sec:fullSpatial}

In this section we give a simple translation from the
first-order logic with spatial conjunction and inductive
definitions \cite[Chapter
4]{Immerman98DescriptiveComplexity} to second-order logic.
This gives an upper bound on the expressive power of
first-order logic with spatial conjunction and inductive
definitions.

Consider first-order logic extended with the spatial
conjunction $\spand$ and the least-fixpoint operator.  The
syntax of the least-fixpoint operator is
\begin{equation*}
  (\lfp\, P,x_1,\ldots,x_n. F)(y_1,\ldots,y_n)
\end{equation*}
where $F$ is a formula that may contain new free variables
$P,x_1,\ldots,x_n$.  The meaning of the least-fixpoint
operator is that the relation which is the least fixpoint of
the monotonic transformation on predicates
\begin{equation*}
  (\lambda x_1,\ldots,x_n.P(x_1,\ldots,x_n)) \mapsto (\lambda x_1,\ldots,x_n.F)
\end{equation*}
holds for $y_1,\ldots,y_n$.  To ensure the monotonicity
of the transformation on predicates, we require
that $P$ occurs only positively in $F$.

\begin{figure}
  \begin{equation*}
    \begin{array}{l}
      \cala = \{ A_1, \ldots, A_n \} \mnl
      \calf = \{ f_1, \ldots, f_m \} \\
      \\
      \tr{F' \spand F''} = \exists 
      \begin{array}[t]{l}
        A'_1,\ldots,A'_n, f'_1,\ldots,f'_m, \mnl
        A''_1,\ldots,A''_n, f''_1,\ldots,f''_m.\ \btr{F' \spand F''} \\
      \end{array} \\
      \btr{F' \spand F''} = \\
      \begin{array}{l}
        \bigwedge\limits_{i=1}^n (\splitOne{A_i}{A'_i}{A''_i}) \land
        \bigwedge\limits_{i=1}^m (\splitTwo{f_i}{f'_i}{f''_i})\ \land \mnl
        \tr{F'}[A_i := A'_i]_{i=1}^n [f_i:=f'_i]_{i=1}^m \ \land \mnl
        \tr{F''}[A_i := A''_i]_{i=1}^n [f_i:=f''_i]_{i=1}^m \mnl
      \end{array} \\
      \splitOne{A}{A'}{A''} \equiv \forall x.\, 
      \begin{array}[t]{l}
        (A(x) \miff (A'(x) \lor A''(x)))\ \land \mnls
        \lnot (A'(x) \land A''(x)) \mnl
      \end{array} \\
      \splitTwo{f}{f'}{f''} \equiv \forall x\, y.\, 
      \begin{array}[t]{l}
        (f(x,y) \miff (f'(x,y) \lor f''(x,y)))\ \land \mnls
        \lnot (f'(x,y) \land f''(x,y)) \mnl
      \end{array} \\
      \\
      \tr{(\lfp\, P,x_1,\ldots,x_n. F)(y_1,\ldots,y_n)} = \\
      \quad \forall P.\, (\forall x_1,\ldots,x_n.\, (F \miff P(x_1,\ldots,x_n))) \implies P(y_1,\ldots,y_n)
    \end{array}
  \end{equation*}
  \caption{Translation of Spatial Conjunction and Inductive Definitions
    into Second-Order Logic\label{fig:spatialToSOL}}
\end{figure}

Figure~\ref{fig:spatialToSOL} presents the translation from
first-order logic extended with spatial conjunction
and least-fixpoint operator to second-order logic.  The
translation directly mimics the semantics of $\spand$ and
$\lfp$.

In second-order logic, the relations in $L = \cala \cup
\calf$ become free variables.  

To translate $\spand$, use second-order quantification to
assert the existence of new unary and binary relations that
partition the relations in $L$ into relations in $L'$ and
$L''$.  Then perform a syntactic replacement of relations in
$L$ with the corresponding relations in $L'$ for the first
formula, and with the corresponding relations in $L''$ for
the second formula.

Translating $\lfp$ is also straightforward.  The property
that $P$ is a fixpoint of $F$ is easily expressible.  To
encode that $y_1,\ldots,y_n$ hold for the {\em least}
fixpoint of $F$, we state that $y_1,\ldots,y_n$ hold for all
fixpoints of $F$, using universal second-order
quantification over $P$.

We also note that the translation of $\spand$ in
Figure~\ref{fig:spatialToSOL} uses only existential
second-order quantification, which points to another class
of formulas where spatial conjunction can be eliminated if
we are only concerned with satisfiability.  Namely, if $F'$
and $F''$ are first-order formulas (without $\spand$ or
$\lfp$), then $F' \spand F''$ is satisfiable iff the
first-order formula $\btr{F' \spand F''}$ in the extended
language is satisfiable.  As a slight generalization, define
the following class of ``interesting'' formulas:
\begin{enumerate}
\item a first-order formula $F$ is an interesting formula;
\item if $F_1$ and $F_2$ are interesting formulas, so is 
  $F_1 \spand F_2$;
\item if $F_1$ and $F_2$ are interesting formulas, so is
  $F_1 \lor F_2$
\end{enumerate}
The satisfiability of each interesting formula is equivalent
to the satisfiability of the corresponding first-order
formula in an extended vocabulary.  In particular, the
satisfiability of the class of formulas formed starting from
formulas in two-variable logic with counting and applying
only $\lor$ and $\spand$ is decidable.



\section{Further Related Work}
\label{sec:related}

Records have been studied in the context of functional and
object-oriented programming languages
\cite{JonesJones99LightweightExtensibleRecordsHaskell,
  Remy89TypecheckingRecordsVariantsNaturalExtensionML,
  Remy92TypingRecordConcatenationFree,
  Wand91TypeInferenceRecordConcatenation,
  HarperPierce91RecordCalculusConcatenation,
  CardelliMitchell94OperationsRecords,
  CheritonWolf87ExtensionsMultiModuleRecords,
  Pottier03ConstraintBasedPresentationGeneralizationRows,
  NaraschewskiWenzel98ObjectOrientedVerificationRecordSubtyping}.
The main difference between existing record notations and
our system is that the interpretation of a record in our
system is a predicate on an object, where an object is
linked to other objects forming a graph, as opposed to being
a type that denotes a value (with values typically
representable as finite trees).  Our view is appropriate for
programming languages such as Java and ML that can
manipulate structures using destructive updates.  Our
generalizations allow the developers to express both
incoming and outgoing references of objects, and allow
the developers to express typestate changes.

We have developed role logic to provide a foundation for
role analysis \cite{KuncakETAL01LanguageRoleSpecifications,
  KuncakETAL01RolesTechRep, Kuncak01DesigningRoleAnalysis,
  KuncakETAL02RoleAnalysis}.  We have subsequently studied a
simplification of role analysis constraints and showed a
characterization of such constraints using formulas
\cite{KuncakRinard02TypestateCheckingRegularGraphConstraints,
  KuncakRinard03ExistentialHeapAbstractionEntailment}.  
Multifields and multislots are present already in
\cite[Section 8.1]{KuncakETAL01RolesTechRep}.  In this
section we have shown that role logic provides a unifying
framework for all these constraints and goes beyond them in
1) being closed under the fundamental boolean logical
operations, and, 2) being closed under spatial conjunction
for an interesting class of formulas.  The
view of roles as predicates is equivalent to the view of
roles as sets and works well in the presence of data abstraction
\cite{LamETAL03OnModularPluggableAnalysesUsingSetInterfaces,
  LamETAL04GeneralizedTypestateCheckingUsingSets}.

The parametric analysis based on there-valued logic was
introduced in \cite{SagivETAL99Parametric,
  SagivETAL02Parametric}.  Other approaches to verifying shape invariants
include \cite{FradetMetayer97ShapeTypes, FradetMetayer98StructuredGamma,
Moeller01PALE, ChaseETAL90AnalysisPointersStructures,
HummelETAL94GeneralDataDependence, GhiyaHendren98PuttingPointerAnalysisWork}.
A decidable logic for expressing connectivity properties of
the heap was presented in
\cite{BenediktETAL99LogicForLinked}.
We use spatial conjunction from separation logic that has
been used for reasoning about the heap
\cite{IshtiaqOHearn01BIAssertionLanguage,
  Reynolds00IntuitionisticReasoningMutable,
  Reynolds02SeparationLogic,
  CalcagnoETAL00SemanticAnalysisPointerAliasings,
  CalcagnoETAL02DecidingValiditySpatialLogicTrees}.
Description logics
\cite{BaaderETAL03DescriptionLogicHandbook,
  Borgida95DescriptionLogicsDataManagement} share many of
the properties of role logic and have been traditionally
applied to knowledge bases.
\cite{Calvanese96FiniteModelReasoningDescriptionLogics,
  Calvanese96UnrestrictedFiniteModelReasoning} present
doubly-exponential deterministic algorithms for reasoning
about the satisfiability of expressive description logics
over all structures and over finite structures.  The
decidability of two-variable logic with counting $C^2$ was
shown in
\cite{GraedelETAL97TwoVariableLogicCountingDecidable},
whereas \cite{PacholskiETAL00ComplexityResultsFirstOrderTwo}
establishes the $\NEXPTIME$-complexity of the satisfiability
problem for the fragment $C^2_1$ with counting up to one.



\section{Conclusions}
\label{sec:conclusions}

We have shown how to add notation for records to
two-variable role logic while preserving its decidability.
The resulting notation supports a generalization of
traditional records with record specifications that are
closed under all boolean operations as well as record
concatenation, allow the description of typestate properties,
support inverse records, and capture the distinction between
open and closed records.  We believe that such an expressive
and decidable notation is useful as an annotation language
used with program analyses and type systems.



\vspace*{1em}
\smartparagraph{Acknowledgements}
We thank the participants of the Dagstuhl Seminar 03101
``Reasoning about Shape'' for useful discussions on
separation logic and shape analysis.


\bibliographystyle{plain}
\bibliography{pnew}

\appendix
\section{Appendix: Correctness of Spatial Conjunction Elimination}

\smartparagraph{Proposition~\ref{prop:qfreeToCAT}}
  Every quantifier-free formula $F$ such that $\FV{F}
  \subseteq \{ x_1,\ldots,x_n \}$ is equivalent to a
  disjunction of CAT formulas $C$ such that $\FV{C} = \{
  x_1,\ldots,x_n \}$.
\begin{proof}
  Let $F$ be a quantifier-free formula
  and $\FV{F} \subseteq \{ x_1,\ldots,x_n\}$.
  Transform $F$ to disjunctive normal form $F'$.
  Let $C$ be a conjunction in $F'$.  If $C$ contains
  a literal and its negation, then $C$ is contradictory
  and we eliminate $C$ from $F'$.  Assume all conjunctions
  are non-contradictory, and let $C$ be one conjunction.
  If there exists an atomic formula $F_A$ in variables
  $\{ x_1,\ldots,x_n \}$ such that $F_A \notin C$ and
  $(\lnot F_A) \notin C$, then replace $C$
  with the disjunction 
  \begin{equationc}
    (C \land F_A) \lor (C \land \lnot F_A)
  \end{equationc}
  By repeating this process, we obtain a disjunction of CAT
  formulas.
\end{proof}

\smartparagraph{Lemma~\ref{lemma:equalingCAT}}
  Every CAT formula $F$ is either contradictory, or is
  equivalent to an EQCAT formula $F'$ such that $\FV{F'} =
  \FV{F}$.
\begin{proof}
  Let $F$ be a CAT formula.  If $x_i \neq x_i$ occurs in
  $F$, then $F$ is contradictory.  If $x_i = x_j$ occurs in
  $F$ for $i \not \equiv j$, then in all conjuncts other
  than $x_i = x_j$ replace all occurrences of $x_i$ with
  $x_j$.  Repeat this process as long as it is possible.
  Suppose that the resulting formula was not established to
  be contradictory.  Let $y_1,\ldots,y_m$ be variables that
  occur only on the left-hand side of some equality $y_j =
  x_{i_j}$.  Removing all equalities of the form $y_j = y_j$
  yields an EQCAT formula.
\end{proof}

\smartparagraph{Proposition~\ref{prop:qfreeToEQCAT}}
  Every quantifier-free formula $F$ such that $\FV{F}
  \subseteq \{ x_1,\ldots, x_n \}$ can be written as a
  disjunction of EQCAT formulas $C$ such that $\FV{C} = \{
  x_1,\ldots,x_n \}$.
\begin{proof}
  Let $F$ be a quantifier-free formula such that $\FV{F}
  \subseteq \{ x_1,\ldots, x_n \}$.  Using
  Proposition~\ref{prop:qfreeToCAT}, transform $F$ to
  disjunction of CAT formulas $F_1$.  Then, for each
  conjunct $C$ of $F_1$ apply Lemma~\ref{lemma:equalingCAT}
  to transform $C$ to an EQCAT formula.
\end{proof}

\smartparagraph{Proposition~\ref{prop:depthOneNF}}
  Let $F$ be a formula of such that $F$ has quantifier depth
  at most one, $F$ has counting degree at most $k$, and
  $\FV{F} \subseteq \{ x_1,\ldots,x_n \}$.  Then $F$ is
  equivalent to a disjunction of $k$-counting-star formulas
  $F_C$ where $\FV{F_C} = \{ x_1,\ldots, x_n \}$.
\begin{proof}
  Let $F$ be a formula of such that $F$ has quantifier depth
  at most one, $F$ has counting degree at most $k$, and
  $\FV{F} \subseteq \{ x_1,\ldots,x_n \}$.  Then $F$ is a
  boolean combination of 1) atomic formulas and 2) formulas of the
  form $\cexists{s}{z}{F'}$ where $F'$ is quantifier-free
  and $\FV{F'} = \{ z,x_1,\ldots,x_n \}$.  Because $z$ is a
  bound variable, rename it to $x$ in each formula $F'$.
  Let $F_1$ be the result of transforming this boolean
  combination to disjunctive normal form.  Consider a
  disjunct $C$ of $F_1$.  As in the proof of
  Proposition~\ref{prop:qfreeToEQCAT}, and treating quantified formulas
  as atomic syntactic entities, 
  transform $C$ into disjunction of formulas of the form
  \begin{equation*}
    \bigwedge\limits_{j=1}^m y_j = w_{i_j} \ \land \ F \ \land
    \bigwedge\limits_{F' \in S} (\cexists{\beta(F')}{x}{F'})^{\alpha(F')}
  \end{equation*}
  where $\beta(F') \in C_{k+1}$, $\alpha(F') \in \{ 0, 1 \}$
  for $F' \in S$, and where $\bigwedge_{j=1}^m y_j = w_{i_j} 
  \land F$ is an EQCAT formula with
  $y_1,\ldots,y_m,w_1,\ldots,w_p$ distinct variables such
  that $\{ y_1,\ldots, y_m, w_1, \ldots, w_p \} = \{
  x_1,\ldots,x_n \}$, and $\FV{F'} \subseteq \{
  x,x_1,\ldots,x_n \}$ for $F' \in S$.  Here $S$ is the set
  of formulas of the form $\cexists{\beta(F')}{x}{F'}$
  that end up conjoined with the EQCAT formula as the result
  of transformation to normal form.
  By replacing each
  $y_j$ with $w_{i_j}$ in each $F'$, enforce that
  $\FV{F'} \subseteq \{ x, w_1,\ldots, w_p \}$.  Using
  Proposition~\ref{prop:qfreeToEQCAT}, transform each $F'$ to a
  disjunction of EQCAT formulas.  By applying the equivalences
  \begin{equation*} 
    \begin{array}{l}
      \existsgeq{k_1}{x}{\bigvee\limits_{i=1}^q B_i} \ssim
      \bigvee\limits_{\sum\limits_{j=1}^q l_j = k_1} \ \
      \bigwedge\limits_{i=1}^q \existsgeq{l_i}{x}{B_i} 
      \mnl
      \existseq{k_1}{x}{\bigvee\limits_{i=1}^q B_i} \ssim
      \bigvee\limits_{\sum\limits_{j=1}^q l_j = k_1} \ \
      \bigwedge\limits_{i=1}^q \existseq{l_i}{x}{B_i} 
    \end{array}
  \end{equation*}
  for $B_1,\ldots,B_q$ mutually exclusive,
  and propagating the disjunction to the top level,
  ensure that every $F'$ is an EQCAT formula.  Then
  transform each term
  $(\cexists{\beta(F')}{x}{F'})^{\alpha(F')}$ into positive
  boolean combination of formulas of one of the forms
  $\existseq{i}{x}{F'}$ for $0 \leq i \leq k$ and
  $\existsgeq{k+1}{x}{F'}$, using the properties
  \begin{equation*}
    \begin{array}{rcl}
      \lnot \existsgeq{k_1}{x}{F'} & \ssim & \bigvee\limits_{i=0}^{k_1{-}1} \existseq{i}{x}{F'} \mnl
      \lnot \existseq{k_1}{x}{F'} & \ssim & \bigvee\limits_{i \in \{0,\ldots,k\}\setminus \{k_1\}} \existseq{i}{x}{F'} \ \lor 
                                     \existsgeq{k{+}1}{x}{F'}
    \end{array}
  \end{equation*}
  Next ensure that each $F'$ is not merely an EQCAT, but
  in fact a GCCAT such that $F' \in \exts{F}{x}$, as follows.
  
  Suppose that $F'$ contains a literal $L_1$ complementary to some
  literal occurring in GCCAT formula $F$.  If $L_1$ occurs in
  $\existseq{i}{x}{F'}$ for $i > 0$ or in
  $\existsgeq{k{+}1}{x}{F'}$, then the entire conjunct is
  contradictory and we eliminate it.  If $L_1$ occurs in
  $\existseq{0}{x}{F'}$, then $\existseq{0}{x}{F'}$ is
  implied by $F$, so eliminate it.
  Assume that $F'$ has no literals complementary to
  literals in $F$.  Then $F'$ contains $w_i \neq
  w_j$ for all $i \not\equiv j$.  Next ensure that $x \neq w_i$
  is a conjunct for $1 \leq i \leq p$, as follows.  
  Suppose that $F'$ contains the conjunct
  $x=w_i$ for some $1 \leq i \leq p$.  
  
  There is clearly at most one interpretation of $x$ that is
  equal to interpretation of $w_i$, so if $\beta(F') \in \{
  2, 3, \ldots, k, \mkplus{(k+1)} \}$ then $F$ and $F'$ are
  contradictory and the entire conjunction is $\boolFalse$,
  so assume $\beta(F') \in \{ 0,1 \}$.  For the same reason,
  $\existseq{1}{x}{F'}$ is equivalent to $\exists x.F'$, so
  if $\beta(F')=1$, then replace $x$ with $w_i$ in $F'$
  giving a GCCAT formula $F''$ such that $\FV{F''}=\FV{F}$.
  By definition of GCCAT formulas, either $F$ and $F''$ are
  equivalent, so $F \land (\exists x.F'') \ssim F$, or $F$
  and $F''$ are contradictory, and the entire conjunction is
  $\boolFalse$.

  Assume therefore that $x \neq w_i$ occurs in $F'$ for all
  $1 \leq i \leq p$.  This means that $F'$ is a GCCAT formula.
  Because $\FV{F'} = \{ x, w_1,\ldots,w_p \}$ and $F'$ does
  not contain a literal complementary to a literal from $F$,
  eliminating from $F'$ atomic formulas that occur in $F$
  yields an element of $\exts{F}{x}$.
  
  To ensure that there exists exactly one conjunct of the
  form $\cexists{s}{x}{F'}$ for each $F' \in \exts{F}{x}$,
  use the fact that the $k+1$ formulas
  $\existseq{i}{x}{F'}$, for $0 \leq i \leq k$, and
  $\existsgeq{k{+}1}{x}{F'}$ form a partition (they are
  mutually exclusive and their disjunction is $\boolTrue$).
\end{proof}

\smartparagraph{Lemma~\ref{lemma:translationCorrectness}}
  If $e$ is an environment for language $L$, $C$ a
  counting star formula in language $L$, and $m \in \{ \{ 1
  \}, \{ 2 \}, \{ 1,2 \} \}$, then $\tr{C}e =
  \str{m}{C}e^m$.
\begin{proof}
  Formula $E$ contains only equalities, so $\tr{E}e$ iff
  $\tr{E}e^m$.  It therefore suffices to show that
  \begin{equation} \label{eqn:spatialTrue}
    \tr{\kstr{F} \spand \xstr{m}{\exists^{s_1} x.F'_1} \spand \ldots 
      \spand \xstr{m}{\exists^{s_k} x.F'_k}}e^m = \boolTrue
  \end{equation}
  iff $\tr{F}e = \boolTrue$ and for all $i$,
  $\tr{\exists^{s_i} x.F'_i}e = \boolTrue$.
  
  $\Rightarrow$): Let~(\ref{eqn:spatialTrue}) hold.  Then
  there exist $e_0, e_1, \ldots, e_k$ such that $\split\ e^m
  [e_0\, e_1\, \ldots e_k]$, $\tr{\kstr{F}}e_0 = \boolTrue$,
  and $\tr{\xstr{m}{\exists^{s_i} x.F'_i}}e_i = \boolTrue$
  for $1 \leq i \leq k$.
  
  We first show $\tr{F}e = \boolTrue$.  Note first that
  $\tr{G_E}e_i = \boolTrue$ for $1 \leq i \leq k$.  Namely,
  because both $\anyNumi{F'}{m}$ and $\oneNumi{F'}{m}$
  entail $G_E$, so does $\xstr{m}{\exists^{s_i} x.F'_i}$, by
  definition of $\xstr{m}{}$ and $\split$.  Therefore, $e_0$
  is the only environment among $e_0,e_1,\ldots,e_k$ that
  may have non-empty relations between the elements
  interpreting $x_1,\ldots,x_n$.  As a result,
  $\tr{F}e^m=\tr{F}e_0$.  But $\tr{F}e_0=\boolTrue$ because
  $\tr{\kstr{F}}e_0 = \boolTrue$.  Therefore
  $\tr{F}e^m=\boolTrue$, and $F$ contains no symbols from
  $L' \setminus L$, so $\tr{F}e=\boolTrue$.
  
  We next show $\tr{\exists^{s_i} x.F'_i}e = \boolTrue$ for
  $1 \leq i \leq k$.  For $s_i = p^{+}$, from
  $\tr{\xstr{m}{\exists^{s_i} x.F'_i}}e_i = \boolTrue$ we
  have that there exist $e_{i,0},e_{i,1},\ldots,e_{i,p}$
  such that 1) $\split\, e_i
  [e_{i,0},e_{i,1},\ldots,e_{i,p}]$, 2)
  $\tr{\anyNumi{F'}{m}}e_{i,0} = \boolTrue$, and 3)
  $\tr{\oneNumi{F'}{m}}e_{i,j} = \boolTrue$ for $1 \leq j
  \leq p$.  Similarly, for $s_i < p$, we have that there
  exist $e_{i,1},\ldots,e_{i,s_i}$ such that 1) $\split\,
  e_i [e_{i,1},\ldots,e_{i,s_i}]$, and 2)
  $\tr{\oneNumi{F'}{m}}e_{i,j} = \boolTrue$ for $1 \leq j
  \leq s_i$.  Note that whenever
  $\tr{\anyNumi{F'}{m}}e_{i,j}$ or
  $\tr{\oneNumi{F'}{m}}e_{i,j}$ holds, we can split
  elements of the domain $D$ into two disjoint sets:
  elements $E_{i,j}$ for which $\allemptyExtension{F}{x}$
  holds, and elements $N_{i,j}$ for which $F' \land
  \Mark{m}{x}$ holds.  If $\tr{\oneNumi{F'}{m}}e_{i,j}$,
  then $|N_{i,j}|=1$, by definition of
  $\tr{\oneNumi{F'}{m}}e_{i,j}$.  Moreover, by definition of
  $\split$ and because $m \neq \emptyset$, we have
  $N_{i_1,j_1} \cap N_{i_2,j_2} = \emptyset$ for
  $\tu{i_1,j_1} \neq \tu{i_2,j_2}$.  Observe that, for a
  given domain element $d \in D$, the atomic type extension
  corresponding to $e^m$ with $x \mapsto d$ is the union of
  atomic type extensions corresponding to each $e_{i,j}$.
  The atomic type extension for $d$ in $e_{i,j}$ is either
  $F' \land \Mark{m}{x}$, or $\allemptyExtension{F}{x}$.
  Therefore, the atomic type extension for $d$ in $e^m$ is
  either $F' \land \Mark{m}{x}$ if $d \in N_{i,j}$ for some
  $i,j$, or $\allemptyExtension{F}{x}$ if for all $i,j$, $d
  \notin N_{i,j}$.  If $N_i = \{ d \mid
  \tr{F'_i}e^m[x \mapsto d] = \boolTrue \}$, then $N_i =
  \biguplus_j N_{i,j}$.  If $s_i=k < p$ then
  $|N_i|=\sum_{j=1}^{s_i}|N_{i,j}| = \sum_{i=j}^{s_i} 1 =
  s_i$, so $\tr{\existseq{k}{x}{F'_i}}e^m = \boolTrue$.  Because
  $\existseq{k}{x}{F'_i}$ is formula in language $L$, we have
  $\tr{\existseq{k}{x}{F'_i}}e = \boolTrue$.  Similarly, if $s_i
  = p^{+}$, then $|N_i|=|N_{i,0}| + \sum_{j=1}^{p}|N_{i,j}|
  = |N_{i,0}| + p \geq p$, so $\tr{\existsgeq{k}{x}{F'_i}}e^m
  = \boolTrue$ and therefore $\tr{\existsgeq{p}{x}{F'_i}}e =
  \boolTrue$.  In both cases, $\tr{\exists^{s_i} x.\, F'_i}e
  = \boolTrue$.
  
  This completes one direction of the implication, we next
  show the converse direction.

  $\Leftarrow$):
  Let $\tr{F}e = \boolTrue$ and for all $i$ where $1 \leq i
  \leq k$, $\tr{\exists^{s_i} x.F'_i}e = \boolTrue$.  We
  construct environments $e_0,e_1,\ldots,e_k$ such that 1)
  $\split\, e^m\, [e_0,e_1,\ldots,e_k]$ 2) $\tr{\kstr{F}}e_0
  = \boolTrue$, and 3) $\tr{\xstr{m}{\exists^{s_i}
      x.F'_i}}e_i = \boolTrue$ for all $i$ where $1 \leq i
  \leq k$.  We construct $e_0,e_1,\ldots,e_k$ by assigning
  the tuples of relations in $e$ to one of the environments
  $e_0,e_1,\ldots,e_k$, as follows.  We only need to decide
  on splitting the tuples $\tu{d_1,\ldots,d_q}$ where all
  but one value $d_1,\ldots,d_q$ are from the set $D_X = \{
  e x_1,\ldots,e x_n \}$, the values of relations on other
  tuples do not affect the truth value of formulas in
  question and can be split arbitrarily.  If
  $\{d_1,\ldots,d_q \} \subseteq D_X$, then we assign the
  tuple to $e_0$, as a result, $\tr{\kstr{F}}e_0 = \boolTrue$.
   If $\{d_1,\ldots,d_q \} \setminus D_X =
  \{ d \}$, then let $i$ be such that $F'_i$ is the unique
  extension of $F$ with the property $\tr{F'_i}e[x \mapsto
  d]=\boolTrue$.  Then assign the tuple
  $\tu{d_1,\ldots,d_q}$ to the environment $e_i$ and also
  assign the values $(e\, B_l)\, d$ for all $l \in m$ to $e_i$.
  Because we assign each relevant tuple to exactly one
  $e_i$, we ensure $\split\, e^m\, [e_0,e_1,\ldots,e_k]$.
  Let $D_E = \{ d \mid \tr{F'_i}e[x \mapsto d] = \boolTrue
  \}$, then also $D_E = \{ d \mid \tr{F'_i}e_i[x \mapsto d]
  = \boolTrue \}$.  Because $\tr{\exists^{s_i} x.F'_i}e =
  \boolTrue$, $|D_E|=s_i$ for $s_i < p$ and $|D_E|\geq p$
  for $s_i = p^{+}$.  Let $s_i < p$.  Then split $e_i$ into
  $e_{i,1},\ldots,e_{i,s_i}$ by assigning exactly one
  element $d \in D_E$ to one $e_{i,j}$.  When assigning an
  element we assign the values of all relations from $L$, as
  well as the relations $B_1$ and $B_2$.  This ensures that
  $\tr{\oneNumi{F'_i}{m}}e_{i,j} = \boolTrue$ for all $1
  \leq i \leq s_i$.  For $s_i = p^{+}$, we split $e_i$ into
  $e_{i,0},e_{i,1},\ldots,e_{i,p}$ by assigning exactly one
  element to each of $e_{i,1},\ldots,e_{i,p}$ and assigning
  the remaining elements to $e_{i,0}$.  In both cases, we
  obtain $\tr{\xstr{m}{\exists^{s_i} x.F'_i}}e_i =
  \boolTrue$.
\end{proof}

\smartparagraph{Lemma~\ref{lemma:rulesSound}}
  If $G_1 \combto G_2$, then $G_2 \implies G_1$ is valid.
\begin{proof}
  We show the claim for each of the rules $(1)$--$(6)$.
  
  Rule $(1)$: Let $T_1 \ispand T_2$ be defined and let
  $\tr{\oneNumi{T_1 \ispand T_2}{1,2}}e = \boolTrue$ for an
  $L'$-environment $e$.  Let $d \in D$ be the unique domain
  element such that $\tr{T_1 \ispand T_2}e[x \mapsto
  d]=\boolTrue$.  Let $e_1$ and $e_2$ be such that $\split\,
  e\, [e_1,e_2]$, $\tr{T_1}e_1[x \mapsto d]=\boolTrue$ and
  $\tr{T_2}e_2[x \mapsto d]=\boolTrue$, and $e_p B_q d =
  \boolTrue$ iff $p=q$ for $p,q \in \{1,2\}$.  In other
  words, $e_1$ and $e_2$ split $e$ by assigning tuples
  validating $T_1$ to $e_1$, tuples validating $T_2$ to
  $e_2$, and by assigning $B_1$ to $e_1$ and $B_2$ to $e_2$
  on the element $d$.  The values of relations $e\, r$
  containing tuples with an element $d' \notin \{ e
  x_1,\ldots, e x_n, d\}$ are all $\boolFalse$, because
  $\tr{\oneNumi{T_1 \ispand T_2}{1,2}}e = \boolTrue$, so we
  let the values of $e_1 r$ and $e_2 r$ for those tuples 
  also be empty.  Then $d$ is the only element outside $\{e
  x_1,\ldots,e x_n\}$ such that $\tr{T_1}e_1[x \mapsto d] =
  \boolTrue$, and $d$ is also the only element outside $\{e
  x_1,\ldots,e x_n\}$ such that $\tr{T_2}e_2[x \mapsto d] =
  \boolTrue$.  As a result, $\tr{\oneNumi{T_1}{1}}e_1 =
  \boolTrue$ and $\tr{\oneNumi{T_2}{2}}e_2 = \boolTrue$, so
  $\tr{\oneNumi{T_1}{1} \spand
    \oneNumi{T_2}{2}}e=\boolTrue$.
  
  To show the claim for rules $(2)$, $(3)$, $(4)$, we
  proceed similarly as for rule $(1)$.
  
  Rule $(2)$: Let $T_1 \ispand T_2$ be defined and let
  $\tr{\oneNumi{T_1 \ispand T_2}{1,2} \spand
    \anyNumi{T_2}{2}}e=\boolTrue$.  Then there are $e'$ and
  $e''$ such that $\split\, e\, [e',e'']$, $\tr{\oneNumi{T_1
      \ispand T_2}{1,2}}e'=\boolTrue$ and
  $\tr{\anyNumi{T_2}{2}}e''=\boolTrue$.  Let $d$ be the
  unique element such that $\tr{T_1 \ispand T_2}e'[x \mapsto
  d]=\boolTrue$, and let $d_1,\ldots,d_k$ be the list of all
  (distinct) elements such that $\tr{\anyNumi{T_2}{2}}e''[x
  \mapsto d_i]=\boolTrue$.  Note that $d \notin \{
  d_1,\ldots,d_k\}$, because $e' B_2 d = \boolTrue$, $e''
  B_2 d_i = \boolTrue$ for all $1 \leq i \leq k$, and
  $\split\, e\, [e',e'']$.  We construct $e_1$ and $e_2$
  such that $\split\, e\, [e_1,e_2]$ as follows.  We assign $B_1$,
  as well as the values of relations that hold according to $T_1$ on
  element $d$ to $e_1$, and we assign $B_2$, as well as the values of
  relations that hold according to $T_2$ on element $d$ to
  $e_2$.  We assign $B_2$ as well as the values of relations that hold
  according to $T_2$ on $d_1,\ldots,d_k$ to $e_2$. The
  values of $B_1$ and the relations on $d_1,\ldots,d_k$ for $e_1$ are
  empty.  For such $e_1$ and $e_2$ we have
  $\tr{\oneNumi{T_1}{1}}e_1=\boolTrue$ and 
  $\tr{\anyNumi{T_2}{2}}e_2=\boolTrue$, so
  $\tr{\oneNumi{T_1}{1} \spand \anyNumi{T_2}{2}}e=\boolTrue$.

  Rule $(3)$ is analogous to rule $(2)$.
  
  Rule $(4)$: Let $T_1 \ispand T_2$ be defined and let
  $\tr{\anyNumi{T_1}{1} \spand \anyNumi{T_2}{2} \spand
    \anyNumi{T_1 \ispand T_2}{1,2}}=\boolTrue$.  Then there
  are $e'$,$e''$, $e'''$ such that $\split\, e\,
  [e',e'',e''']$, $\tr{\anyNumi{T_1}{1}}e'=\boolTrue$,
  $\tr{\anyNumi{T_2}{2}}e''=\boolTrue$, and $\tr{\anyNumi{T_1
      \ispand T_2}{1,2}}e'''=\boolTrue$.  Then there are
  three sets of elements $N'$, $N''$, $N'''$, where $N'$
  contains elements that validate $T_1$ in $e'$, $N''$
  contains elements that validate $T_2$ in $e''$, and $N'''$
  contains elements that validate $T_1 \ispand T_2$ in
  $e'''$.  We have $N' \cap N'''=\emptyset$ and $N'' \cap
  N'''=\emptyset$, whereas $N' \cap N''$ need not be empty.
  Each element $d \notin \{e x_1,\ldots, e x_n\}$
  validates in $e$ either 1) $\allemptyExtension{F}{x}$, if $d
  \notin N' \cup N'' \cup N'''$, or 2) $T_1$, if $d \in N'
  \setminus N''$, or 3) $T_2$, if $d \in N'' \setminus N'$,
  or 4) $T_1 \ispand T_2$, if $d \in (N' \cap N'') \cup
  N'''$.  We construct environments $e_1$,$e_2$,$e_3$ by
  assigning $B_1$ and relations from $T_1$ to elements in $N'
  \setminus N''$ to $e_1$, assigning $B_2$ and elements in
  $N' \setminus N''$ to $e_2$, and splitting relations on
  elements in $(N' \cap N'') \cup N'''$ into those for $T_1$,
  which we assign to $e_1$, and those for $T_2$, which we
  assign to $e_2$.  We then have
  $\tr{\anyNumi{T_1}{1}}e_1 = \boolTrue$ and
  $\tr{\anyNumi{T_2}{2}}e_2 = \boolTrue$, so
  $\tr{\anyNumi{T_1}{1} \spand \anyNumi{T_2}{2}}=\boolTrue$.

  Rules $(5)$, $(6)$: Directly from the definitions of $\empe$
  and $\anyNumi{F'}{m}$ it follows that $\empe \implies
  \anyNumi{F'}{m}$.
\end{proof}

\smartparagraph{Lemma~\ref{lemma:easyDirection}}
  $(\bigvee_{\ERes{C_1}{C_2}{C_3}} C_3) \implies (C_1 \spand
  C_2)$ is a valid formula for every pair of counting star
  formulas $C_1$ and $C_2$.
\begin{proof}
  Let $\tr{\bigvee_{\ERes{C_1}{C_2}{C_3}} C_3}e$ hold for
  some $L$-environment $e$.
  Then $\tr{C_3}e=\boolTrue$ for some $C_3$ such that
  $\str{1}{C_1} \spand \str{2}{C_2} \yields \str{1,2}{C_3}$.
  By Lemma~\ref{lemma:yieldSound},
  $\str{1,2}{C_3} \implies \str{1}{C_1} \spand \str{2}{C_2}$ is valid.
  By Lemma~\ref{lemma:translationCorrectness} and $\tr{C_3}e=\boolTrue$,
  we have $\tr{\str{1,2}{C_3}}e^{1,2}=\boolTrue$.  Therefore,
  $\tr{\str{1}{C_1} \spand \str{2}{C_2}}e^{1,2}=\boolTrue$.
  This means that there are $e_1$ and $e_2$ such that
  $\split\, e^{1,2}\, [e_1,e_2]$, 
  $\tr{\str{1}{C_1}}e_1=\boolTrue$, and
  $\tr{\str{2}{C_2}}e_2=\boolTrue$.
  From Lemma~\ref{lemma:translationCorrectness} we have
  $\tr{C_1}\eproj{e_1}=\boolTrue$, and 
  $\tr{C_2}\eproj{e_2}=\boolTrue$.  From 
  $\split\, e^{1,2}\, [e_1,e_2]$ it follows that
  $\split\, e\, [\eproj{e_1},\eproj{e_2}]$, so
  $\tr{C_1 \spand C_2}e=\boolTrue$.
\end{proof}

\smartparagraph{Lemma~\ref{lemma:difficultDirection}}
  $C_1 \spand C_2 \implies \bigvee_{\ERes{C_1}{C_2}{C_3}} C_3$ 
  is a valid formula for every pair of counting star
  formulas $C_1$ and $C_2$.
\begin{proof}
  Let $\tr{C_1 \spand C_2}e=\boolTrue$ for some
  $L$-environment $e$.  Then there are $e_1$ and $e_2$ such
  that $\split\, e\, [e_1,e_2]$, $\tr{C_1}e_1=\boolTrue$ and
  $\tr{C_2}e_2=\boolTrue$.  By
  Lemma~\ref{lemma:translationCorrectness},
  $\str{1}{C_1}e_1^1=\boolTrue$ and
  $\str{2}{C_2}e_2^2=\boolTrue$.  We construct
  $\str{1,2}{C_3}$ such that $\str{1}{C_1} \spand
  \str{2}{C_2} \yields \str{1,2}{C_3}$ and
  $\tr{C_3}e=\boolTrue$, as follows.
  
  Let $K_1$ be the GCCAT part of $C_1$ and let
  $K_2$ be the GCCAT part of $C_2$.
  Let $D_X = D \setminus \{ e\, x_1,\ldots,e\, x_n\}$.  For each
  $d \in D_X$, let $T_1^d$ be the type extension induced by
  $d$ in $e_1$, that is, let $T_1^d \in \exts{K_1}{x}$ be the
  formula such that $\tr{T_1^d}e^1_1[x\mapsto d]=\boolTrue$.
  Similarly, let $T_2^d \in \exts{K_2}{x}$ be the formula such
  that $\tr{T_2^d}e^2_2[x \mapsto d]=\boolTrue$.  Because
  $\split\, e\, [e_1,e_2]$, the operation $T_1 \ispand T_2$
  is defined and $\tr{T_1 \ispand T_2}e^{1,2}[x \mapsto
  d]=\boolTrue$.  Because $\str{1}{C_1}e_1^1=\boolTrue$,
  with each $d$ we can associate an occurrence $\mu_1(d)$ in
  $\str{1}{C_1}$ of a formula $F_{\mu_1(d)}$ where
  $F_{\mu_1(d)}$ is of the form $\oneNumi{T_1^d}{1}$ or of
  the form $\anyNumi{T_1^d}{1}$, and an environment
  $e_{1,\mu_1(d)}$ such that $\split\, e_1^1\,
  [e_{1,0}, (e_{1,\mu_1(d)})_{\mu(d)}]$, such that 
  $\kstr{K_1}e_{1,0}=\boolTrue$, and such that
  for every $d$,
  $\tr{F_{\mu_1(d)}}e_{1,\mu_1(d)} = \boolTrue$.
  Analogously, for each $d$ we can associate an occurrence
  $\mu_2(d)$ in $\str{2}{C_2}$ of a formula $F_{\mu_2(d)}$
  of the form $\oneNumi{T_2^d}{2}$ or of the form
  $\anyNumi{T_2^d}{2}$, and an environment $e_{2,\mu_2(d)}$
  such that $\split\, e_2^2\, [e_{2,0}, (e_{2,\mu_2(d)})_{\mu_2(d)}]$,
  such that $\kstr{K_2}e_{2,0}=\boolTrue$, 
  and such that for every $d$,
  $\tr{F_{\mu_2(d)}}e_{2,\mu_2(d)} = \boolTrue$.
  
  We compute $C_3$ by first combining $\kstr{K_1}$ and
  $\kstr{K_2}$ into $\kstr{K_1 \kispand K_2}$.
  From $\split\, e\, [e_1,e_2]$ we conclude that
  the operation
  $F_1 \kispand F_2$ is well-defined and that
  $\tr{\kstr{F_1 \kispand F_2}}e^{1,2}_0 = \boolTrue$ where
  $e^{1,2}_0$ is given by $\split\, e^{1,2}_0\, [e_{1,0},e_{2,0}]$.

  We next apply rules
  $(1)$--$(4)$ in Figure~\ref{fig:transRules}, as follows:
  \begin{enumerate}
  \item apply rule $(1)$ once to each pair of occurrences $\mu_1(d)$ and $\mu_2(d)$
    if they are of the form $\oneNumi{T_1^d}{1}$ and
    $\oneNumi{T_2^d}{2}$, respectively; let $\mu(d)$ be the occurrence
    of the resulting formula $F_{\mu(d)} \equiv \oneNumi{T_1^d \ispand T_2^d}{1,2}$;
  \item apply rule $(2)$ once to each pair of occurrences $\mu_1(d)$ and $\mu_2(d)$
    if $\mu_1(d)$ is an occurrence of the form $\oneNumi{T_1^d}{1}$ and
    $\mu_2(d)$ is an occurrence of the form $\anyNumi{T_2^d}{2}$;
    let $\mu(d)$ be the occurrence
    of the formula $F_{\mu(d)} \equiv \oneNumi{T_1^d \ispand T_2^d}{1,2}$ obtained as one of the results;
  \item apply rule $(3)$ once to each pair of occurrences $\mu_1(d)$ and $\mu_2(d)$
    if $\mu_1(d)$ is an occurrence of the form $\anyNumi{T_1^d}{1}$ and
    $\mu_2(d)$ is an occurrence of the form $\oneNumi{T_2^d}{2}$;
    let $\mu(d)$ be the occurrence
    of the formula $F_{\mu(d)} \equiv 
    \oneNumi{T_1^d \ispand T_2^d}{1,2}$ obtained as one of the results;
  \item apply rule $(4)$ once for each pair of occurrences of
    formulas of the form $\anyNumi{T_1^d}{1}$ and $\anyNumi{T_2^d}{2}$;
    for each $d$ such that $\mu_1(d)$ is an occurrence of $\anyNumi{T_1^d}{1}$ and
     $\mu_2(d)$ is an occurrence of $\anyNumi{T_2^d}{2}$,
    let $\mu(d)$ be the occurrence of the resulting formula
    $F_{\mu(d)} \equiv \anyNumi{T_1^d \ispand T_2^d}{1,2}$.
  \end{enumerate}
  Note that no rule is applied twice to a distinct pair of
  occurrences of formulas.  This means that the number of
  applications of rules is uniformly bounded, despite the
  fact that there is no bound on the size of the model $e$.
  In particular, there is no bound on the number of elements
  $d$ covered by a single application of rule $(4)$.  Each
  formula of the form $\oneNumi{T}{1}$ is $F_{\mu_1(d)}$ for
  some $d$ and each formula of the form $\oneNumi{T}{2}$ is
  $F_{\mu_2(d)}$ for some $d$, and all such formulas are
  consumed by applications of rules $(1)$--$(3)$, so the
  resulting formula has no subformulas of the form
  $\oneNumi{T}{1}$ or $\oneNumi{T}{2}$.  After applying
  rules $(1)$--$(4)$, apply rules $(5)$ and $(6)$ to all
  applicable formulas.  The resulting formula $F_R$ has no
  occurrences of $\anyNumi{T}{1}$ or $\anyNumi{T}{2}$
  either, it contains only occurrences of formulas of forms
  $\oneNumi{T}{1,2}$ and $\anyNumi{T}{1,2}$.
  
  For each of the finitely many occurrences $\mu(d)$ in
  $F_R$ we construct $e^{1,2}_{\mu(d)}$, splitting $e^{1,2}$
  into the environment $e^{1,2}_0$ defined above, and the
  environments $e^{1,2}_{\mu(d)}$, by assigning the type
  extension of $d$ in $e^{1,2}$ to $e^{1,2}_{\mu(d)}$.  By
  construction, $\split\, e^{1,2}\, [e^{1,2}_0,
  (e^{1,2}_{\mu(d)})_{\mu(d)}]$.  To show
  $\tr{F_R}e^{1,2}=\boolTrue$, it suffices to show
  \begin{equation} \label{eqn:resConjunct}
    \tr{F_c}e^{1,2}_{c} = \boolTrue
  \end{equation}
  for every occurrence $c = \mu(d_0)$.  Fix an occurrence $c$,
  and let $\delta = \{ d \mid \mu(d)=c \}$.  By definition
  of $e^{1,2}_c$, the type extension induced by each $d \in
  \delta$ in $e^{1,2}_c$ is $T_1^d \ispand T_2^d$, and the
  type extension of each $d \in D_X \setminus \delta$ is an
  empty extension.  Therefore, $\tr{\anyNumi{T_1^d \ispand
      T_2^d}{1,2}}e^{1,2}_c = \boolTrue$.  If $F_c \equiv \anyNumi{T_1^d
    \ispand T_2^d}{1,2}$ then the equation (\ref{eqn:resConjunct})
  already holds.  If $F_c \equiv \oneNumi{T_1^d \ispand
    T_2^d}{1,2}$, then $F_c$ was generated by one of the
  rules $(1)$--$(3)$, which means that $\delta$ is a
  singleton set.  Namely, if $F_c$ was generated by rules
  $(1)$ or $(2)$, then there is exactly one $d$ such that
  $\mu_1(d)=c$, namely $d_0$, and similarly if $F_c$ was
  generated by rule $(3)$, then there is exactly one $d$
  such that $\mu_2(d)=c$, again $d_0$.  In both cases,
  $\delta = \{ d_0 \}$, so $d_0$ is the unique $d$ with type
  extension $T_1^d \ispand T_2^d$, which means that
  $\tr{\oneNumi{T_1^d \ispand
      T_2^d}{1,2}}e^{1,2}_c=\boolTrue$ and the equation
  (\ref{eqn:resConjunct}) holds.

  We finally apply idempotence to ensure that no
  $\anyNumi{T}{m}$ occurs more than once.  The resulting
  formula $F'_R$ is equivalent to $F_R$, so
  $\tr{F'_R}e^{1,2}=\boolTrue$, $F'_R$ is of the form
  $\str{1,2}{C_3}$, and $\str{1}{C_1} \spand \str{2}{C_2}
  \yields \str{1,2}{C_3}$.  From $\str{1,2}{C_3}$ we recover
  $C_3$ using the inverse of the translation in
  Figure~\ref{fig:translationToSpatial}.  By
  Lemma~\ref{lemma:translationCorrectness} we have
  $\tr{C_3}e=\boolTrue$, completing the proof.
\end{proof}



\end{document}